\documentclass[prb,aps,amssymb,twocolumn,superscriptaddress,notitlepage]{revtex4-1}
\usepackage{amsmath}
\usepackage{amssymb}
\usepackage{amsthm}
\usepackage{amsfonts}
\usepackage{mathrsfs}
\usepackage{listings}
\usepackage{enumerate}
\usepackage{latexsym}
\usepackage{psfrag}
\usepackage{bm}
\usepackage{graphicx}
\usepackage[caption=false]{subfig}
\usepackage{blkarray}
\usepackage{array}
\usepackage{color}
\usepackage[normalem]{ulem}
\usepackage{hyperref}

\def\blue{\color{black}}

\def\kdp{\mathbf{k}\cdot\mathbf{p}}

\newcolumntype{L}{>{$}l<{$}}

\newtheorem{defn}{Definition}
\newtheorem*{defn*}{Definition}

\newcommand{\half}{\frac{1}{2}}
\newcommand\Tstrut{\rule{0pt}{2.6ex}}  

\begin{document}
\title{Band Connectivity for Topological Quantum Chemistry: Band Structures As A Graph Theory Problem}
\author{Barry Bradlyn}
\thanks{These authors contributed equally to the preparation of this work.}
\affiliation{Princeton Center for Theoretical Science, Princeton University, Princeton, New Jersey 08544, USA}
\author{L. Elcoro}
\thanks{These authors contributed equally to the preparation of this work.}
\affiliation{Department of Condensed Matter Physics, University of the Basque Country UPV/EHU, Apartado 644, 48080 Bilbao, Spain}
\author{M.~G. Vergniory}
\thanks{These authors contributed equally to the preparation of this work.}
\affiliation{Donostia International Physics Center, P. Manuel de Lardizabal 4, 20018 Donostia-San Sebasti\'{a}n, Spain}
\affiliation{Department of Applied Physics II, University of the Basque Country UPV/EHU, Apartado 644, 48080 Bilbao, Spain}
\affiliation{Ikerbasque, Basque Foundation for Science, 48013 Bilbao, Spain}
\author{Jennifer Cano}
\affiliation{Princeton Center for Theoretical Science, Princeton University, Princeton, New Jersey 08544, USA}
\author{Zhijun Wang}
\affiliation{Department of Physics, Princeton University, Princeton, New Jersey 08544, USA}
\author{C. Felser}
\affiliation{Max Planck Institute for Chemical Physics of Solids, 01187 Dresden, Germany}
\author{M.~I.~Aroyo}
\affiliation{Department of Condensed Matter Physics, University of the Basque Country UPV/EHU, Apartado 644, 48080 Bilbao, Spain}
\author{B. Andrei Bernevig}
\thanks{Permanent Address: Department of Physics, Princeton University, Princeton, New Jersey 08544, USA }
\affiliation{Department of Physics, Princeton University, Princeton, New Jersey 08544, USA}
\affiliation{Donostia International Physics Center, P. Manuel de Lardizabal 4, 20018 Donostia-San Sebasti\'{a}n, Spain}
\affiliation{Laboratoire Pierre Aigrain, Ecole Normale Sup\'{e}rieure-PSL Research University, CNRS, Universit\'{e} Pierre et Marie Curie-Sorbonne Universit\'{e}s, Universit\'{e} Paris Diderot-Sorbonne Paris Cit\'{e}, 24 rue Lhomond, 75231 Paris Cedex 05, France}
\affiliation{Sorbonne Universit\'{e}s, UPMC Univ Paris 06, UMR 7589, LPTHE, F-75005, Paris, France}
\affiliation{LPTMS, CNRS (UMR 8626), Universit\'e Paris-Saclay, 15 rue Georges Cl\'emenceau,\\ 91405 Orsay, France}
\date{\today}
\begin{abstract}
The conventional theory of solids is well suited to describing band structures locally near isolated points in momentum space, but struggles to capture the full, global picture necessary for understanding topological phenomena. In part of a recent paper [Bradlyn et al., Nature {\bf 547}, 298--305 (2017)], we have introduced the way to overcome this difficulty by formulating the problem of sewing together many disconnected local $\kdp$ band structures across the Brillouin zone in terms of graph theory. In the current manuscript we give the details of our full theoretical construction.  We show that crystal symmetries strongly constrain the allowed connectivities of energy bands, and we employ graph-theoretic techniques such as graph connectivity to enumerate all the solutions to these constraints. The tools of graph theory allow us to  identify disconnected groups of bands in these solutions, and so identify topologically distinct insulating phases. 
\end{abstract}
\maketitle
\section{Introduction}\label{sec:intro}
The fundamental {\blue assumption} of the textbook approach to the theory of solids is that one reaps  enormous benefit by trading the dependence of wavefunctions on real-space position for a dependence on crystal momentum $\mathbf{k}$. Through Bloch's theorem, this has led to the development of the $\kdp$ expansion, allowing for tractable approximations to material Hamiltonians near isolated points in momentum space. This approach found extraordinary successes throughout the twentieth century in the understanding and predicting the behavior of semiconductors, metals (and their Fermi surfaces), and insulators.  

However, an essential shortcoming of this approach is that it is local in momentum space. This renders obscure the \emph{global} properties of wavefunctions in momentum space. It is for precisely this reason that topological insulators appear foreign in the $\kdp$ approach to solid state physics. Isolated $\kdp$ Hamiltonians at various points in the Brillouin zone (BZ) are \emph{a priori} completely independent, and so extracting global (topological) data from such a local momentum space picture seems hopeless at first sight. Such global data could, if existent, be used to characterize all possible band structures (materials) in nature. In particular, one useful outcome of such a characterization would be that it contains both topologically trivial and topologically non-trivial band-structures.

In a part of a recent paper\cite{NaturePaper}, we have introduced, amongst several other concepts, a new way of providing such a global classification of band structures using graph theory, thereby updating band theory with its last missing ingredient. In this paper, we fill in the mathematical details necessary for a full and complete description of our theory.  We map the problem at hand -- patching together isolated $\kdp$ expansions into consistent global band structures -- to a tractable problem in \emph{graph theory}. We first re-interpret the $\kdp$ expansion as a representation subducing technique, and identify representations with the nodes of a "band" graph.  We then show that band structures consistent with the symmetries of a crystal can be put in one-to-one correspondence with graphs that incorporate the symmetries as constraints on how different edges can be joined together. We show that there are often multiple distinct allowed graphs, with different numbers of connected components corresponding to (disconnected) groups of energy bands. In tandem with results of Refs.~\onlinecite{NaturePaper,EBRTheoryPaper}, we will argue that these different connectivities correspond to topologically distinct phases. Thus, we will show that the physics of topological insulators can be captured by the connectivity of band structures, without the need to invoke the tools of differential geometry. This generalizes the known {\blue eigenvalue- and $K$-theory based approaches} to computing topological indices\cite{Fu2007,Teo08,Turner2010,Turner2012,Fang2012,Freed2013,Morimoto2013,Jadaun2013,Chiu2013,Chiu2014,Liu2014,Lu2014,Aris2014,Shiozaki2014,Po2017,Watanabe15,Shiozaki2015,Watanabe16,kanegraphs,po2016filling,Shiozaki2017}, unifying them all through the lens of graph theory.

The structure of this paper is as follows. First, in Sec.~\ref{sec:kdp} we review how space group symmetries constrain Hamiltonians locally in momentum space, recasting $\kdp$ perturbation theory as a phenomenological approach to local band structure. In striving to make this paper self-contained, we will also review here the basic properties of crystal symmetry groups and their representations. Next, in Sec.~\ref{sec:global}, we approach the problem of extending locally defined expansions of energy bands to global band structures. We show that crystal symmetries place strong constraints on how different local expansions are allowed to connect to one another. This allows us to map the problem of band connectivity to one of constructing certain types of abstract graphs. In Sec.~\ref{sec:graphs}, we will show how to construct the solutions to this graph problem. The graph mapping allows us to leverage the tools of spectral graph theory, which gives an immediate way to decompose graphs into disconnected components. Reinterpreting this in terms of band structures, we will show how to enumerate the disconnected groups of bands allowed in a global band structure. Finally, in Sec.~\ref{sec:breps}, we will show how the graph approach, along with the theory of band representations developed in Refs.~\onlinecite{Zak1980,Zak1981,Zak1982,Bacry1988,Michel2001,NaturePaper,EBRTheoryPaper}, allows us to reinterpret topological phase transitions as connectivity transitions in graphs. We outline how this mapping, while being an elegant solution of the problem of finding global band theory structure, also allows us to design a constructive algorithm for finding new topological insulators, as was briefly demonstrated in Ref.~\onlinecite{NaturePaper}.

\section{The $\mathbf{k}\cdot\mathbf{p}$ Method: Momentum-space locality}\label{sec:kdp}

The $\mathbf{k}\cdot\mathbf{p}$ method has been enormously successful in the study of band structures near special points in the Brillouin zone, especially when one is only interested in a few isolated bands at a time. Traditionally, this method has been employed to fit effective masses and coupling constants to real band structures. Here, however, we will focus on the constraints placed on $\mathbf{k}\cdot\mathbf{p}$ by symmetries and re-interpret the $\kdp$ method into a problem of subducing representations away from a high-symmetry point onto high-symmetry lines. Unlike most textbook treatments\cite{Kittel87,Grosso}, our discussion will apply not only to expansions around the origin of the Brillouin zone (the $\Gamma$ point), but rather to \emph{any and every} high-symmetry $\mathbf{k}$-vector.

\subsection{Symmetric Bloch Hamiltonians}
Let us consider the Bloch Hamiltonian $H(\mathbf{k})$ for a crystal with the symmetries of some space group $G$. The symmetry of the crystal implies that $H(\mathbf{k})$ transforms {\blue as a scalar under} some representation $\Delta$ of $G$, such that for all $g\in G$
\begin{equation}
\Delta(g)H(\mathbf{k})\Delta(g^{-1})=H(g\mathbf{k}).\label{eq:hamsymmetry}
\end{equation}
We will be interested in the behavior of $H$ near some high-symmetry point ${\mathbf{k}_0}$ in the BZ. {\blue Let us introduce the little group $G_{\mathbf{k}_0}$ of $\mathbf{k}_0$, defined as

\begin{equation}
G_{\mathbf{k}_0}\equiv\{\{R|\mathbf{t}\}\in G | R\mathbf{k}_0=\mathbf{k}_0+\mathbf{K}\},
\end{equation}
where $\mathbf{K}$ is a reciprocal lattice vector}. This is the subgroup $G_{{\mathbf{k}_0}}\subset G$ of the space group which leaves ${\mathbf{k}_0}$ invariant up to a reciprocal lattice vector. {\blue The little group plays a privileged role in constraining the $\kdp$ theory near $\mathbf{k}_0$.} Restricted to the little group, the representation $\Delta$ subduces to
\begin{equation}
\Delta_{{\mathbf{k}_0}}\equiv \Delta\downarrow G_{{\mathbf{k}_0}}=\bigoplus_i\rho_i,\label{eq:sgsubducedrep}
\end{equation}
where the $\rho_i$ are irreducible representations acting on vector spaces $V_i$. Using Eq.~(\ref{eq:hamsymmetry}) we see that
\begin{equation}
[\Delta_{{\mathbf{k}_0}}(g),H({\mathbf{k}_0})]=0
\end{equation}
for all $g\in G_{\mathbf{k}_0}$: the matrix representatives of the elements of $G_{\mathbf{k}_0}$ commute with $H({\mathbf{k}_0})$. Schur's Lemma then tells us\cite{Fulton2004} that the matrix $H({\mathbf{k}_0})$ in the representation space of $\Delta_{\mathbf{k}_0}$ is a sum of constants $\epsilon_i$ for each irrep $\rho_i$ in the decomposition Eq.~(\ref{eq:sgsubducedrep}),
\begin{equation}
H({\mathbf{k}_0})=\bigoplus_i \epsilon_i P_{\rho_i},\label{eq:pointham}
\end{equation}
where $P_{\rho_i}$ is the projector onto the representation space $V_i$ of $\rho_i$. {\blue For those representations which occur with multiplicity larger than one, Schur's Lemma implies only that the Hamiltonian is block diagonal; we assume here that we have carried out any additional diagonalization needed to put the Hamiltonian in the form Eq.~(\ref{eq:pointham}).} Thus, electronic states at ${\mathbf{k}_0}$ come in degenerate sets of energies $\epsilon_i$ and degeneracy given by the dimension of the representations $\rho_i$.

We thus see that at the high-symmetry point ${\mathbf{k}_0}$ the structure of the Hamiltonian is almost trivial. Things become more interesting, however, when we look slightly away from ${\mathbf{k}_0}$. Here the representatives of the little group $G_{{\mathbf{k}_0}}$ no longer commute with $H(\mathbf{k})$, although they still place strong contraints. Writing $\mathbf{k}={\mathbf{k}_0}+\delta\mathbf{k}$, we may expand $H(\mathbf{k})$ in powers of $\delta\mathbf{k}$
\begin{equation}
H(\mathbf{k})=H({\mathbf{k}_0})+\delta k^\mu H^{(1)}_\mu+\delta k^\mu\delta k^\nu H^{(2)}_{\mu\nu}+\dots,\label{eq:kdotpexpansion}
\end{equation}
{\blue where we have introduced the matrix-valued expansion coefficients $H^{(i)}$.} We use $\mu,\nu=1,2,\dots, D$ to index the {\blue primitive basis} directions in the BZ, and repeated indices are summed; $H({\mathbf{k}_0})$ is given by Eq.~(\ref{eq:pointham}). We now observe that, if the energy spacings $\epsilon_i-\epsilon_j$ are the largest scales in the problem, which is always the case for small $\delta k$, the matrices $H^{(\alpha)}$ appearing in the expansion Eq.~(\ref{eq:kdotpexpansion}) are approximately block diagonal in the carrier space $\oplus_iV_i$ of irreducible representations of $G_{{\mathbf{k}_0}}$. More precisely, we may write in perturbation theory
\begin{equation}
H^{(\alpha)}_{\mu_1\mu_2\dots\mu_n}=\bigoplus_i H^{(\alpha),i}_{\mu_1\mu_2\dots\mu_n}P_{\rho_i}+\mathcal{O}(\delta\epsilon^{-1}),
\end{equation}
where $\delta\epsilon=\mathrm{min}_{i\neq j}(|\epsilon_i-\epsilon_j|)$. Note that although each $H^{(\alpha),i}$ acts only within the space $V_i$, they are not diagonal matrices within this space; rather from Eq.~(\ref{eq:hamsymmetry}) we deduce that for each $g\in G_{{\mathbf{k}_0}}$
\begin{align}
\rho_i(g)H^{(\alpha),i}_{\mu_1\mu_2\dots\mu_n}&\rho_i(g^{-1})\delta{k}_{\mu_1}\delta{k}_{\mu_2}\cdots\delta{k}_{\mu_n}=\nonumber \\
&H^{(\alpha),i}_{\mu_1\mu_2\dots\mu_n}(g\delta\mathbf{k})_{\mu_1}(g\delta\mathbf{k})_{\mu_2}\cdots(g\delta\mathbf{k})_{\mu_n}
\end{align}
The practical consequence of this, of course, is that near to ${\mathbf{k}_0}$, we may truncate the expansion Eq.~(\ref{eq:kdotpexpansion}), ignore mixing of degenerate groups of bands at ${\mathbf{k}_0}$, and for small but nonzero $\delta\mathbf{k}$ faithfully reproduce the spectrum and eigenstates of the full Hamiltonian $H(\mathbf{k})$. {\blue Away from $\mathbf{k}_0$, the points $\mathbf{k}_0+\delta\mathbf{k}$ have their own little groups $G_{\mathbf{k}_0+\delta{k}}\subset G_{\mathbf{k}_0}$. Bands and eigenstates away from $\mathbf{k}_0$ thus transform under representations of $G_{\mathbf{k}_0+\delta\mathbf{k}_0}$ which are subduced\cite{GroupTheoryPaper} from representations of $G_{\mathbf{k}_0}$ (as we will explore further in Sec.~\ref{sec:comprel}).} Because the representations $\rho_i$ of the little group $G_{{\mathbf{k}_0}}$ play a special role in the $\kdp$ method, we shall explore their properties in more detail. We will confine the discussion, however, only to what we will need for writing and patching together $\kdp$ Hamiltonians; a full account of the theory may be found in Ref.~\onlinecite{Cracknell}.   

\subsection{Little Groups and Their Representations}
We now delve into the structure and representation theory of the little groups. Recall that a space group $G$ consists of elements of the form $\{R|\mathbf{d}\}$, where $R$ is a rotation or rotoinversion, and $\mathbf{d}$ is a translation. Each space group contains a lattice generated by $\{E|\mathbf{t}_i\},i=1,\dots,D$ in $D$-dimensions (we use $E$ to denote the identity element). The BZ in momentum space is a unit cell of the reciprocal lattice, with basis vectors $\mathbf{g}_i$ satisfying $\mathbf{g}_i\cdot\mathbf{t}_j=2\pi\delta_{ij}$.

The little group $G_\mathbf{k}$ of a point $\mathbf{k}$ consists of all elements $\{R|\mathbf{d}\}\in G$ in the space group $G$ such that $R\mathbf{k}\equiv\mathbf{k}$, where equivalence is defined up to a reciprocal lattice {\blue vector}.
Note in particular that the little group contains all pure lattice translations $\{E|\mathbf{t}\}$, since real-space (direct space) translations are local in momentum space and do not change $\mathbf{k}$, and so $G_\mathbf{k}$ is itself a space group. The point group of the little group ${G}_\mathbf{k}$ 
is known as the little co-group, denoted by $\overline{G}_\mathbf{k}$.

Turning to representations, we recall the (somewhat obvious) fact that the momentum $\mathbf{k}$ is the quantum number for the action of translations on Bloch states. As such, when looking at Bloch Hamiltonians, we only consider those irreducible representations $\rho$ of $G_\mathbf{k}$ such that
\begin{equation}
\rho(\{E|\mathbf{t}\})=e^{-i\mathbf{k}\cdot\mathbf{t}}{\blue \mathbb{I}},\label{eq:smallrep}
\end{equation}
{\blue where $\mathbb{I}$ is the identity matrix.} Representations of this type are conventionally called \emph{allowed representations}\cite{Cracknell}. However, since we will never be speaking of ``disallowed'' representations {\blue (those in which lattice translations are not represented by phases, for instance via augmented matrices\cite{ITCsymmetry}. Such representations occur in the theory of band representations\cite{Zak1982,Evarestov1997,NaturePaper,EBRTheoryPaper})}, we shall omit the word ``allowed'' without risk of confusion.

At the $\Gamma$ point in the BZ, $\mathbf{k}=0$, and so all representations $\rho^\Gamma_i$ of $G_{\Gamma}\approx G$ satisfy $\rho^\Gamma_i(\{E|\mathbf{t}\})=\mathbb{I}$. Because of this, we may identify representations of $G_{\Gamma}$ with representations of the little co-group $\overline{G}_\Gamma$; representations of $G_\Gamma$ are {\blue the same for all space groups $G$ that share the same point group $\bar{G}$, regardless of whether they are symmorphic or non-symmorphic}. 

Away from the $\Gamma$ point, however, the situation is more interesting. The phases in Eq.~(\ref{eq:smallrep}) do not {\blue in general} vanish at high-symmetry points other than $\Gamma$, and so the representations of $G_\mathbf{k}$, and in particular their degeneracies, are sensitive to the symmorphicity of the space group $G$. In particular, screw rotations and glide reflections in non-symmorphic groups yield pure translations when raised to appropriate powers. While these are represented as the identity at $\Gamma$, they yield non-trivial phases at other $\mathbf{k}$ points. As such, {\blue in non-symmorphic groups} and for {\blue $\mathbf{k}\neq\mathbf{0}$ lying on a screw axis, in a glide plane, or at the boundary of the BZ}, the representations of the little group $G_{\mathbf{k}}$ are \emph{projective} representations of the little co-group $\overline{G}_\mathbf{k}$. This has profound consequences on the $\mathbf{k}\cdot\mathbf{p}$ Hamiltonian, as we shall see in the example in Sec.~\ref{sec:KLexample}.

{\blue Additionally, we must distinguish between the transformation properties of spinless and spin-$1/2$ (or colloquially, ``spinful'') particles.}
For spinless particles, we know that a rotation by $2\pi$ should leave the wavefunction invariant, while for spin-$1/2$ particles, a rotation by $2\pi$ multiplies the wavefunction by $-1$. 
{\blue To accomodate this in the theory of little group representations, we introduce the \emph{double space groups} and their representations\cite{Cracknell,GroupTheoryPaper}. The double groups are central extensions of the space group, obtained by adjoining an element $\overline{E}$ which commutes with all elements of $G_\mathbf{k}$. 
This element signifies a $2\pi$ rotation, and so we extend the groups so that every $n$-fold rotation, when raised to the $n$-th power, yields $\overline{E}$.}
Representations of the double groups are termed single- or double-valued based on their value on $\overline{E}$. In particular, we have
\begin{equation}
\rho(\{\overline{E}R|\mathbf{d}\})=\begin{cases}
\rho(\{R|\mathbf{d}\}), &\rho\;\mathrm{is\;single{\textrm -}valued} \\
-\rho(\{R|\mathbf{d}\}), &\rho\;\mathrm{is\;double{\textrm -}valued}.
\end{cases}
\end{equation}
{\blue For the remainder of this work, we will focus solely on the double groups; we will follow the accepted convention\cite{cdml,ssc}} of distinguishing double-valued representations with the use of an overbar. 

Finally, we will often be interested in systems with time-reversal (TR) symmetry. Space groups with time reversal symmetry contain an additional antiunitary element $T$, which commutes with all other elements of the space group, leaves real-space (direct space) invariant, and maps $\mathbf{k}$ to $-\mathbf{k}$. With TR, it is important to distinguish between three different types of $\mathbf{k}$-vectors and little groups. 

First, there are those vectors ${\mathbf{k}_0}$ such that $-{\mathbf{k}_0}\equiv{\mathbf{k}_0}$. In this case, $T\in G_{{\mathbf{k}_0}}$. In spinless representations $\rho$ of $G_{{\mathbf{k}_0}}$, we have {\blue $\rho(T^2)=\rho(T)^2=\mathbb{I}$}, while for spinful representations $\overline{\rho}(T)^2=-\mathbb{I}$.

Second, it may be the case that $-{\mathbf{k}_0}\not\equiv{\mathbf{k}_0}$, but that some element $g\in G$ maps ${\mathbf{k}_0}$ to $-{\mathbf{k}_0}$. In this case, both $gT$ and $g^2$ are in the little group $G_{{\mathbf{k}_0}}$. We then have for spinless representations $\rho(gT)^2=\rho(g^2)$, and similarly $\overline{\rho}(gT)^2=-\overline{\rho}(g^2)$ for double-valued representations.

Lastly, it may be the case that $-{\mathbf{k}_0}\not\equiv{\mathbf{k}_0}$, and no element of the space group relates ${\mathbf{k}_0}$ and $-{\mathbf{k}_0}$ in this case, TR does not affect the local representation properties of the little group $G_{{\mathbf{k}_0}}$. It will, however, relate globally representations at ${\mathbf{k}_0}$ and $-{\mathbf{k}_0}$ vectors in a $\kdp$ expansions.

\subsection{Example: Cubic Crystals}\label{sec:KLexample}
As an example of the above ideas, let us examine the $\kdp$ Hamiltonian for two closely related cubic crystals: the symmorphic group $I432$ ($211$), {\blue and the related non-symmorphic group $I4_132$ ($214$), which share the same point group}. {\blue We take as a basis for the BCC lattice the three vectors
\begin{align}
\mathbf{e}_1&=\frac{1}{2}\left(-\mathbf{\hat{x}}+\mathbf{\hat{y}}+\mathbf{\hat{z}}\right)\nonumber\\
\mathbf{e}_2&=\frac{1}{2}\left(\mathbf{\hat{x}}-\mathbf{\hat{y}}+\mathbf{\hat{z}}\right)\nonumber\\
\mathbf{e}_3&=\frac{1}{2}\left(\mathbf{\hat{x}}+\mathbf{\hat{y}}-\mathbf{\hat{z}}\right).\label{eq:bccvecs}
\end{align}
Both} space groups have point group $O$ ($432$), the octahedral group, consisting of all orientation-preserving symmetries of the cube. However, in $I4_132$, the fourfold rotation $C_{4x}$ about the $x$-axis is embedded into the space group as a screw rotation $\{C_{4x}|00\half\}$, a difference which has profound consequences on the properties of little group representations {\blue [here and throughout, we give translations with respect to the primitive lattice as defined in Eq.~\ref{eq:bccvecs}. However, when convenient, we will label rotations by their Cartesian axes (for instance $C_{4x}$)]}. To see this, let us examine simple four-band $\kdp$ expansions about both the $\Gamma$ point, located at the origin of the BZ, and the $P$ point at the corner of the BZ, with reduced coordinates $\frac{1}{4}\mathbf{g}_1+\frac{1}{4}\mathbf{g}_2+\frac{1}{4}\mathbf{g}_3\equiv(\frac{1}{4}\frac{1}{4}\frac{1}{4})$ in terms of the reciprocal lattice vectors $\mathbf{g}_i$, with Cartesian components
\begin{equation}
\mathbf{g}_1=\frac{2\pi}{a}(0,1,1),\;\mathbf{g}_2=\frac{2\pi}{a}(1,0,1),\;\mathbf{g}_3=\frac{2\pi}{a}(1,1,0).
\end{equation}
In both of these SGs, the little co-group $\overline{G}_\Gamma$ of the $\Gamma$ point is given by the point group $O$, while the little co-group {\blue $\overline{G}_P$} of the $P$ point is the tetrahedral group $T$, the subgroup of $O$ obtained by removing all fourfold rotations, {\blue as well as removing the twofold rotations along the diagonals}.

Let us first examine the Hamiltonian near the $\Gamma$ point. As remarked above, by Eq.~(\ref{eq:smallrep}), {\blue the representations matrices for symmetry elements corresponding to the same point group operation in the little group of $\Gamma$ are identical for both space groups $I432$ and $I4_132$}, and follows trivially from the representation theory of the little co-group $\overline{G}_\Gamma\approx O$. The spinful irreps of the octahedral group are obtained {\blue directly from subduction of the half-integer spin-$J$ irreps of $SU(2)$}, and the representation spaces are spanned by $\{\ell=0,J=\half\},\{\ell=1,J=\half\}$ or $\{\ell=1,J=\frac{3}{2}\}$ basis functions. We focus on four bands transforming in the $J=\frac{3}{2}$ representation, which is conventionally denoted $\overline{\Gamma}_8$ in both space groups. Enforcing the symmetries on the four-band $\kdp$ Hamiltonian in this representation yields in both cases\cite{Bradlyn2016}
\begin{widetext}
\begin{equation}
H^{(211)}(\delta\mathbf{k})=H^{(214)}(\delta\mathbf{k})\approx \epsilon_0\mathbb{I}+\left(\begin{array}{cccc}
a\delta k_z & 0 &-\frac{a+3b}{4}\delta k_+ & \frac{\sqrt{3}}{4}(a-b)\delta k_- \\
0 & b\delta k_z & \frac{\sqrt{3}}{4}(a-b)\delta k_- & -\frac{3a+b}{4}\delta k_+ \\
-\frac{a+3b}{4}\delta k_- & \frac{\sqrt{3}}{4}(a-b)\delta k_+ & -a \delta k_z & 0 \\
\frac{\sqrt{3}}{4}(a-b)\delta k_+ & -\frac{`3a+b}{4}\delta k_- & 0 & -b\delta k_z
\end{array}\right),
\end{equation} 
\end{widetext}
in terms of two real-valued phenomenological parameters $a$ and $b$, and where we have defined $\delta k_\pm=\delta k_x\pm i\delta k_y$. This describes a fourfold degeneracy at $\Gamma$ that disperses linearly, and corresponds to the ``spin-3/2'' fermion introduced in Ref.~\onlinecite{Bradlyn2016}.

Looking next at the $P$ point, we find a very different situation. Without TR, the little co-group $\overline{G}_P\approx T$ of $P$ in both cases is generated by a $C_{3}$ rotation about the cubic body-diagonal, and a twofold rotation $C_{2x}$. With TR, in both cases the little co-group is augmented by the antiunitary operation $\mathcal{A}=C_{2,\hat{x}-\hat{y}}T$, which maps the $P$ point to itself. The little co-group $\overline{G}_P$ both with and without $\mathcal{A}$ has only two-dimensional spinful representations, since it contains the Pauli group of $C_{2x},C_{2y},C_{2z},E$, and their squares {\blue (a fourfold rotation is needed to make the spin-$3/2$ representation irreducible.)}. {\blue For the following, we will consider a TR-invariant system, so that the little co-group contains $\mathcal{A}$.}
For SG $I432$ the representations of the little co-group determine uniquely the representations of the little group acting on the $\kdp$ Hamiltonian. Focusing still on $J=3/2$ states, we find that they transform in the
\emph{reducible} representation of the little group conventionally denoted $\overline{P}_6\oplus\overline{P}_7$. The representation space of $\overline{P}_6$ is spanned by the states $|\frac{3}{2},\frac{1}{2}\rangle-i|\frac{3}{2},-\frac{3}{2}\rangle$ and $|\frac{3}{2},\frac{3}{2}\rangle+i|\frac{3}{2},-\frac{1}{2}\rangle$. Similarly, the $\overline{P}_7$ representation is spanned by states $|J,m\rangle$ of the form $|\frac{3}{2},\frac{1}{2}\rangle+i|\frac{3}{2},-\frac{3}{2}\rangle$ and $|\frac{3}{2},\frac{3}{2}\rangle-i|\frac{3}{2},-\frac{1}{2}\rangle$.
Note that the antiunitary element $\mathcal{A}$ does not couple these two representations. Enforcing these symmetries on the expansion Eq.~(\ref{eq:kdotpexpansion}) about ${\mathbf{k}_0}=\mathbf{k}_P$ yields the block-diagonal Hamiltonian
\begin{equation}
H^{(211)}(\mathbf{k}_P+\delta\mathbf{k})\approx\left(\begin{array}{cc}
\epsilon_1+v_1\delta\mathbf{k}\cdot\mathbf{\sigma} &\mathbf{0} \\
\mathbf{0} & \epsilon_2+v_2\delta\mathbf{k}\cdot\mathbf{\sigma}
\end{array}\right),
\end{equation}
where $\mathbf{\sigma}$ is the vector of Pauli matrices. This describes a pair of twofold degeneracies at $\mathbf{K}_P$ with different energies that disperse linearly away from $P$, i.e~a pair of Weyl fermions at energies $\epsilon_1$ and $\epsilon_2$.

Turning now to SG $I4_132$, we find a completely different phenomenon. Although the little co-group $\overline{G}_P$ is unchanged, the non-symmorphic character of the space group {\blue manifests itself} in the little group $G_P$: this group is generated by the rotation $\{C_3|000\}$ about the cubic body diagonal, and also the twofold \emph{screw} rotation $\{C_{2x}|\overline{\half}\half0\}${\blue ; these two generators fully determine the extension of $\overline{G}_P$ by the group of lattice translations}. In any representation $\rho$ of $G_P$ we must have{\blue \footnote{we remind the reader that the translation part of group operations is given with respect to the primitive basis.}}
\begin{equation}
\rho(\{C_{2x}|\overline{\half}\half0\}^2)=\rho(\{\overline{E}|011\})=e^{-i\pi}\rho(\{\overline{E}|000\}).
\end{equation}
{\blue We recognize the three-band lower-right block of the Hamiltonian as the ``spin-$1$'' Weyl Hamiltonian of Ref.~\onlinecite{Bradlyn2016}.} From this equation, we see that for spin-$1/2$ systems, the representative of the $C_{2x}$ screw in the little group $G_P$ squares to $+\mathbb{I}$. In effect, double-valued representations of the little group $G_P$ are {\blue isomorphic to} \emph{single-valued} representations of the little co-group $\bar{G}_P$. Thus, as was discussed extensively in Sec.~I of the Supplementary Material of Ref.~\onlinecite{Bradlyn2016}, bands at the $P$ point in SG $I4_132$ transform in the sum of a one dimensional representation $\overline{P}_4$ and a three dimensional representation $\overline{P}^{(\mathrm{NS})}_7$ of $G_P$; we have added the superscript ``$(\mathrm{NS})$'' to distinguish this representation from the $2D$ representation in space group $I432$. For the $\kdp$ expansion for the four bands considered here, this yields to linear order
\begin{equation}
H^{(214)}(\mathbf{k}_P+\delta\mathbf{k})\approx\left(\begin{array}{cccc}
\epsilon_1 &0 &0 &0 \\
0& \epsilon_2 &a\delta k_x &a^*\delta k_y \\
0&a^*\delta k_x & \epsilon_2 &a\delta k_z \\
0& a^*\delta k_z& a\delta k_y & \epsilon_2 \\
\end{array}\right).
\end{equation}
We can see from this example that the degeneracies and structure of local Bloch Hamiltonians depends strongly on the symmorphicity of the crystal symmetry group. This is also expected to be the case at points in between $\Gamma$ and $P$. At this point, however, we have no way of connecting the expanded $\kdp$ Hamiltonians at $\Gamma$ and $P$ to form a consistent global band structure. In the next section, we will develop the tools necessary to do so.

\section{Patching together $\mathbf{k}\cdot\mathbf{p}$ Hamiltonians: Global band structures}\label{sec:global}
In order to patch together the local band structures obtained from the $\kdp$ expansions at various points, we will make use of the constraints imposed by crystal symmetry. Having just reformulated the traditional $\kdp$ theory in terms of group representations, we will be able to show how the compatibility between little group representations along high-symmetry lines and planes strongly constrains the allowed connections between bands near different $\mathbf{k}$ points. In order to solve these constraints, we will map the problem to a problem in \emph{graph theory}.

\subsection{Compatibility relations}\label{sec:comprel}
To begin, we examine the symmetry constraints on energy bands at different $\mathbf{k}$-vectors $\mathbf{k}_1$ and $\mathbf{k}_t$, where the little group $G_{\mathbf{k}_t}$ is a proper subgroup of the little group $G_{\mathbf{k}_1}$. This occurs, for instance, when $\mathbf{k}_1$ is a high-symmetry point in the BZ, and $\mathbf{k}_t$ is a point on a high-symmetry line {\blue with endpoint $\mathbf{k}_1$}. For instance, in the example of SG $I432$ in Sec.~\ref{sec:KLexample}, we could take $\mathbf{k}_1$ as the $\Gamma$ point, and $\mathbf{k}_t=(k_t,k_t,k_t)$, which lies on the high-symmetry line $\Lambda$, with little co-group generated only by the $C_3$ rotation about the cubic body diagonal. In all such cases, the relation $G_{\mathbf{k}_t}\subset G_{\mathbf{k}_1}$ implies that each irrep $\rho$ of $G_{\mathbf{k}_1}\supset G_{\mathbf{k}_t}$ restricts to (subduces) {\blue a direct sum of (generally more than one) irreps $\bigoplus_i\sigma_i$} of $G_{\mathbf{k}_t}$, which we denote by
\begin{equation}
\rho\downarrow G_{\mathbf{k}_t}\approx\bigoplus_i\sigma_i. \label{eq:comprel}
\end{equation} 
The restriction is obtained by removing from $\rho$ the matrices for those elements not in $G_{\mathbf{k}_t}$, and viewing the group representation formed by the remaining matrices as a (in general reducible) representation of $G_{\mathbf{k}_t}$.

At this point in the argument, we remark that there exists a whole manifold (line, or plane) of $\mathbf{k}$-vectors with little group $G_{\mathbf{k}_t}$, which we denote $\{\mathbf{k}_t\}$. In the example of SG $I432$ for instance, all $\mathbf{k}$ points of the form $\{\mathbf{k}_t\}=\{\frac{t}{4}(\mathbf{g}_1+\mathbf{g}_2+\mathbf{g}_3),t\in[0,1]\}$ have the same little group $G_\Lambda$ (the endpoints have the larger little groups $G_\Gamma$ and $G_P$, both of which contain $G_\Lambda$). In these situations, the restriction $\rho\downarrow G_{\mathbf{k}_t}$ of little group representations holds along the \emph{entire} manifold of points $\{\mathbf{k}_t\}$. This has two main consequences for $\kdp$ Hamiltonians. First, we know that the $\kdp$ Hamiltonian $H(\mathbf{k}_t+\delta\mathbf{k})$ {\blue around point $\mathbf{k}_t$} transforms {\blue according to} the same representation (and therefore takes the same form) along the entire manifold $\{\mathbf{k}_t\}$. Second, we know that if we focus on a Hamiltonian near $\mathbf{k}_1$ transforming in some representation $\rho$ of $G_{\mathbf{k}_1}$, we know that for $\mathbf{k}_1+\delta\mathbf{k}_\parallel\in\{\mathbf{k}_t\}$, the Hamiltonian $H(\mathbf{k}_1+\delta\mathbf{k}_\parallel)$ must transform in the restricted representation $\rho\downarrow G_{\mathbf{k}_t}$. Thus, the \emph{compatibility relations} Eq.~(\ref{eq:comprel}) between representations of the little groups $G_{\mathbf{k}_1}$ and $G_{\mathbf{k}_t}$ constrains the representation space of Bloch functions that may appear in a band structure. 

\subsection{Global band structures}\label{sec:globalbs}
The compatibility of $\kdp$ band structures gives us our first clue on how to piece spectra at different $\mathbf{k}$ points together: along lower-symmetry manifolds emanating from high-symmetry $\mathbf{k}$-vectors, the representations of electronic states appearing in the Hamiltonian must be compatible. We can complete the picture by noting that the {\blue closure of manifolds} $\{\mathbf{k}_t\}$ of points with little group $G_{\mathbf{k}_t}$ contain more than one high-symmetry $\mathbf{k}$-vector $\mathbf{k}_1$ and $\mathbf{k}_2$, {\blue namely the points at the boundary of $\{k_\mathbf{t}\}$.} For example, in our cubic crystal, with $G_{\mathbf{k}_t}=G_\Lambda$, we have that $\mathbf{k}_1$ is the $\Gamma$ point, and $\mathbf{k}_2$ is the $P$ point. Not only do we have $G_{\mathbf{k}_t}\subset G_{\mathbf{k}_1}$, but also $G_{\mathbf{k}_t}\subset G_{\mathbf{k}_2}$. This allows us to patch together $\kdp$ expansions about the points $\mathbf{k}_1$ and $\mathbf{k}_2$, along the manifold $\{\mathbf{k}_t\}$. If we have a Hamiltonian near $\mathbf{k}_1$ that transforms {\blue according to} a representation $\rho_1$ of $G_{\mathbf{k}_1}$, and a Hamiltonian near $\mathbf{k}_2$ that transforms {\blue according to} a representations $\rho_2$ of $G_{\mathbf{k}_2}$, then these can be consistently connected only if the representations
\begin{equation}
\rho_1\downarrow G_{\mathbf{k}_t}\approx\rho_2\downarrow G_{\mathbf{k}_t}\approx\bigoplus_i\sigma_i
\end{equation}
restricted to the manifold $\{\mathbf{k}_t\}$ are {\blue equivalent}. Furthermore, along the manifold $\{\mathbf{k}_t\}$ we can trace bands carrying each representation $\sigma_i$ from an irrep of $G_{\mathbf{k}_1}$ at $\mathbf{k}_1$ all the way to an irrep of $G_{\mathbf{k}_2}$ at $\mathbf{k}_2$.

\begin{figure}[t]
\subfloat[]{
	\includegraphics[width=0.2\textwidth]{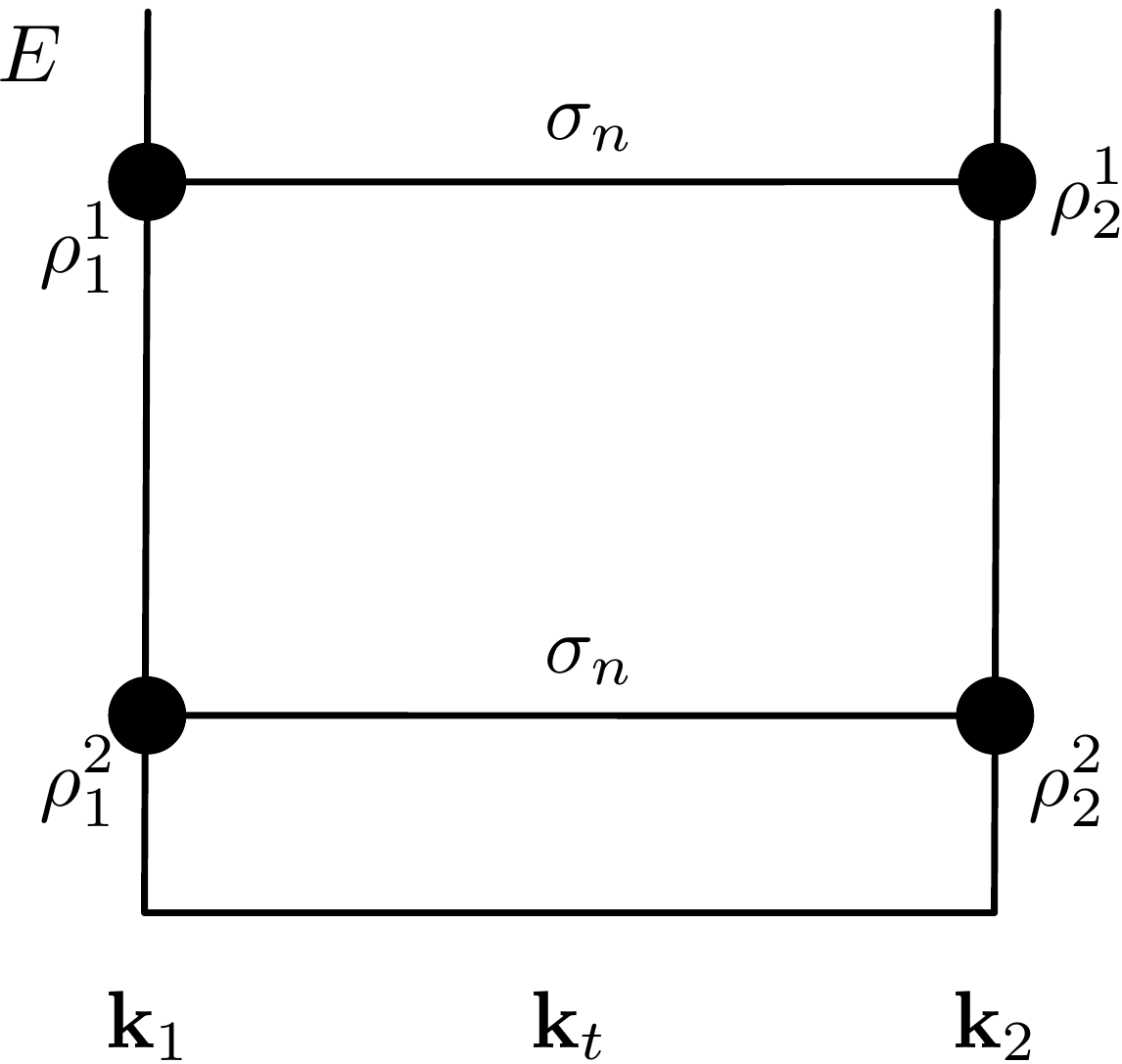}
}
\subfloat[]{
	\includegraphics[width=0.2\textwidth]{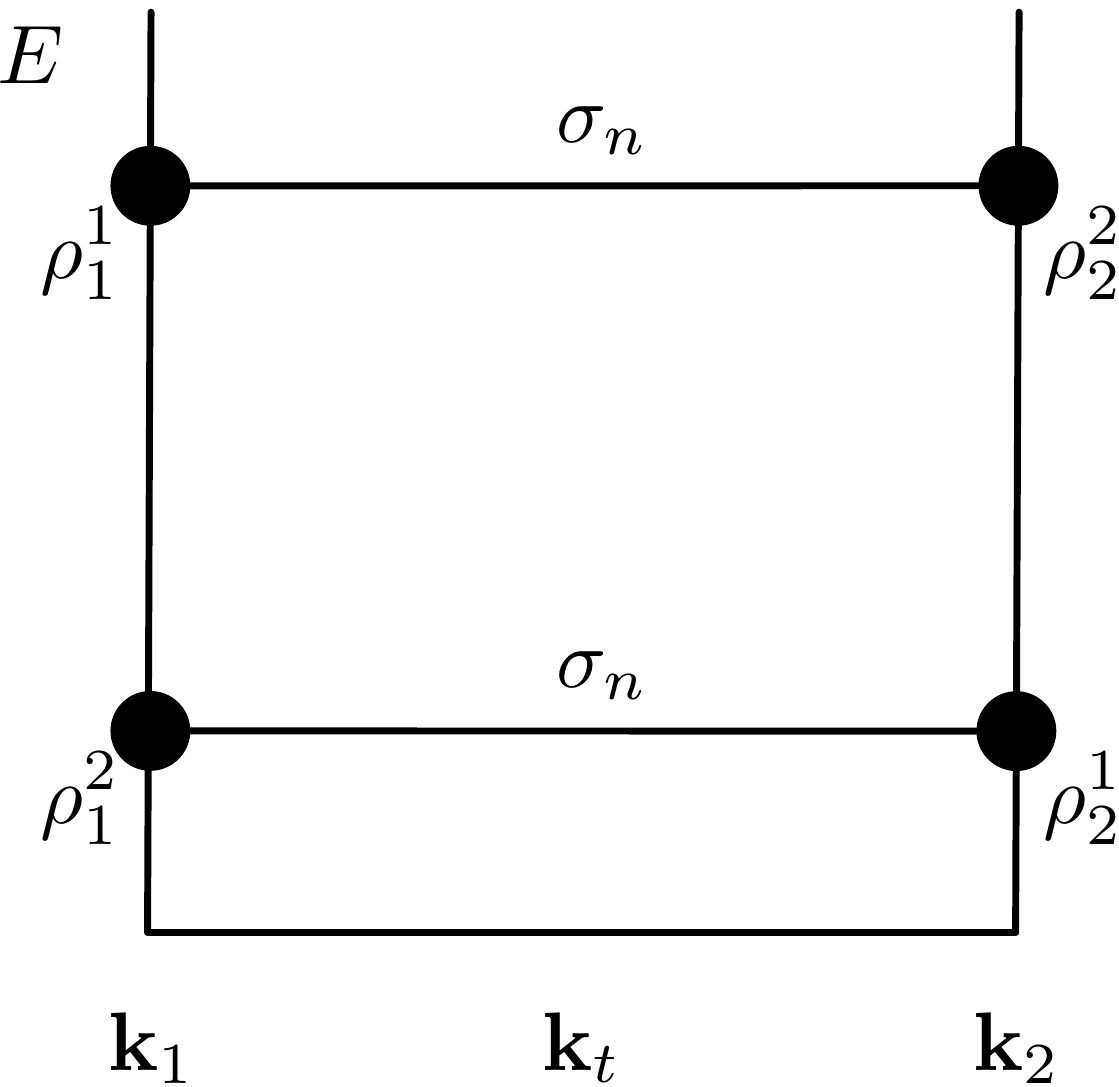}
}
\caption{Two distinct allowed global band structures along the line $\mathbf{k}_1\leftrightarrow\mathbf{k}_t\leftrightarrow\mathbf{k}_2$. $\rho_1^1$ and $\rho_1^2$ are two distinct representations of $G_{\mathbf{k}_1}$, and $\rho_2^1$ and $\rho_2^2$ are two distinct representations of $G_{\mathbf{k}_2}$. All four of these representations subduce the representation $\sigma_n$ of the little group $G_{\mathbf{k}_t}$ of the line $\{\mathbf{k}_t\}$. Because of this, global energy bands may be connected in two ways. In the band structure (a), the states transforming in representation $\rho_1^1$ at $\mathbf{k}_1$ connect to the states transforming in the representation $\rho_2^1$ at $\mathbf{k}_2$. Contrarily, in (b), the states transforming in representation $\rho_1^1$ at $\mathbf{k}_1$ connect to the states transforming in the representation $\rho_2^2$ at $\mathbf{k}_2$.
}\label{fig:distinctbands}
\end{figure}

Extending this logic to \emph{all} $\mathbf{k}$-vectors in the BZ, we can say that patching together $\kdp$ Hamiltonians consistently requires ensuring that {\blue the set of little group representations appearing in the spectrum at each $\mathbf{k}$-vector manifold} are compatible. Even once this is ensured, however, there still may be an ambiguity on how bands join together. This happens when there is a representation $\sigma_n$ that appears in the subductions $\rho_1\downarrow G_{\mathbf{k}_1}$ and $\rho_1\downarrow G_{\mathbf{k}_2}$ with multiplicity $m_n>1$. In this case, multiple bands along the the manifold $\{\mathbf{k}_t\}$ carry the representation $\sigma_n$. If these bands are nondegenerate at both $\mathbf{k}_1$ and $\mathbf{k}_2$, then there are different inequivalent ways of connecting the band structure consistent with the symmetries. We illustrate this schematically in Fig.~\ref{fig:distinctbands}{\blue , where the representation $\sigma_n$ appears with multiplicity $2$, allowing for the two distinct connectivities shown in $(a)$ and $(b)$. More complicated examples quickly arise in other space groups.}

We thus seek a method to find all distinct ways of patching together local band structures to form global spectra, consistent with the compatibility constraints of crystal symmetry. In order to do this efficiently, we will map this to a graph theory problem: vertices in the graph will correspond to irreps of the little group at each symmetry distinct (manifold of) $\mathbf{k}$-vector(s), and two vertices can be joined by an edge only if the corresponding bands connect (implying that the representations are compatible). We will formalize this in the following section, and show how it allows us to fully classify and enumerate allowed band structures.

\section{Constructing connectivity graphs}\label{sec:graphs}
In order to map the problem of finding global band structures to graph theory, we first introduce some basic terminology and concepts in Sec.~\ref{sec:graphreview}. We then define {\blue in Def.~\ref{def:graphs}} the precise mapping {\blue from band structures to ``connectivity graphs''}, and show how global band structures can be constructed by analyzing these graphs one subgraph at a time. We then revisit our cubic crystal example from Sec.~\ref{sec:KLexample}, and show how to patch together the $\kdp$ band structures at the $\Gamma$ and $P$ points in both the symmorphic ($I432$) and non-symmorphic ($I4_132$) cases. Finally, we show how spectral graph theory allows us to analyze the connectivity of global band structures directly in graph-theoretic language. 

\subsection{Review of graph theory}\label{sec:graphreview}

We start by reviewing some basic notions about graphs which will be useful throughout the text. A complete treatment of graph theory can be found in Refs.~\onlinecite{ModernGraphThy,GraphThy}.

\begin{figure}[t]
\includegraphics[width=0.5\textwidth]{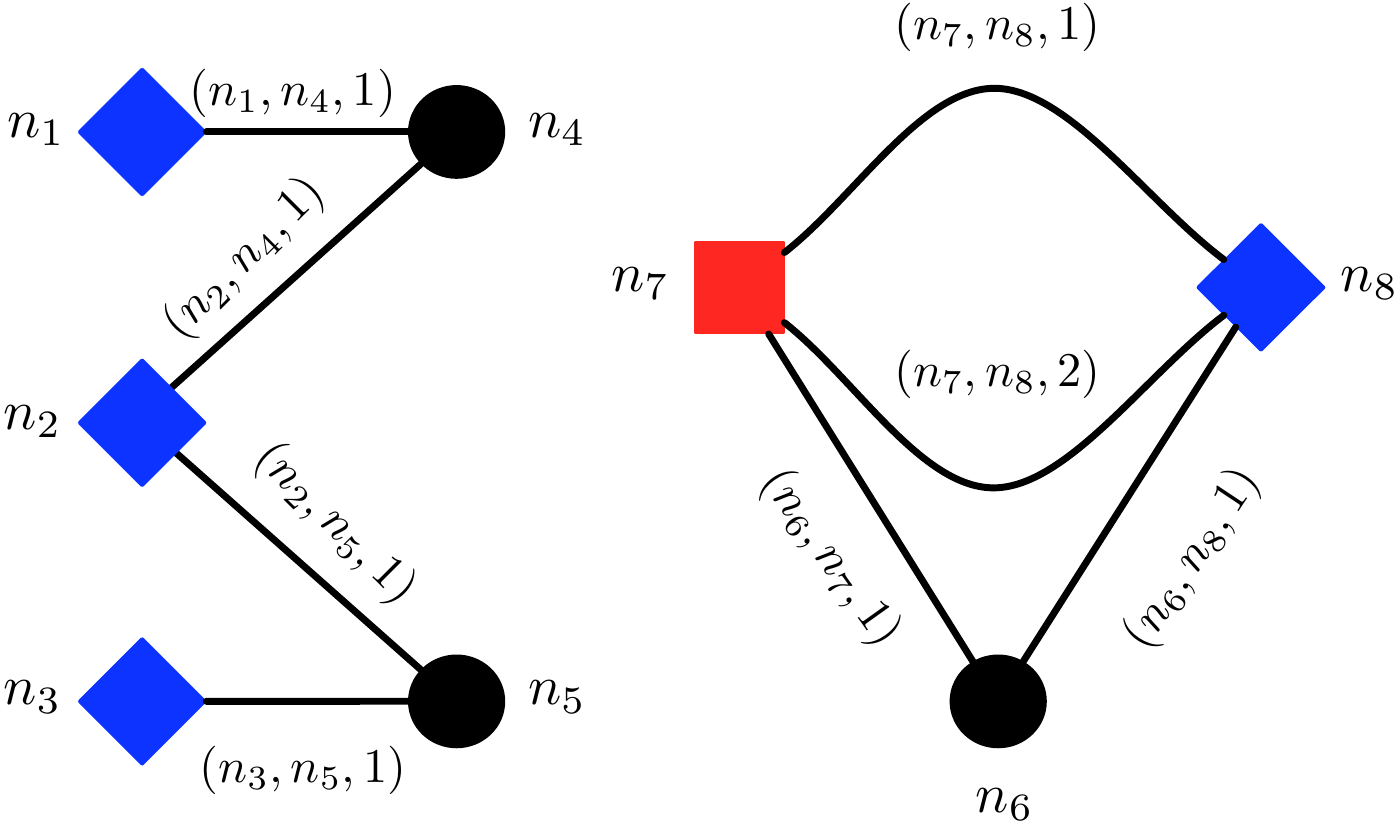}
\caption{Example of a graph. There are eight nodes labelled $n_1$ through $n_8$, indicated by blue diamonds, black circles, and red squares. There are {\blue eight} edges, shown as solid black lines connecting pairs of nodes. Nodes in the graph can be placed (amongst other options) in following three partitions: $\{n_1,n_2,n_3,n_8\}, \{n_4,n_5,n_6\}$, and $\{n_7\}$, since there are no edges connecting nodes within each of these sets; the different colors and symbols for the nodes correspond to this partitoning. This graph has two connected components: the first is the subgraph on the left consisting of the nodes $\{n_1,n_2,n_3,n_4,n_5\}$, and the four edges connected to them. The second connected component consists of the three nodes $\{n_6,n_7,n_8\}$ and the four edges connected to them.}\label{fig:graphdefs}
\end{figure}

A {\bf graph} $\mathcal{G}$ consists of two sets: a set of nodes (or vertices) $N(\mathcal{G})$, and a set of edges $E(\mathcal{G})$. Each edge $e\in E(\mathcal{G})$ connects two distinct nodes $n_1,n_2\in N(\mathcal{G})$. {\blue If two nodes $n_1$  and $n_2$ are connected by $\ell > 0$ edges, we label the edges as} $(n_1,n_2,\alpha)$, where $\alpha\in\{1,\dots \ell_{n_1n_2}\}$ (note that if $\ell_{n_1n_2}>1$, the graph contains $\ell_{n_1n_2}-1$ loops connecting nodes $n_1$ and $n_2$); furthermore, since $n_1$ and $n_2$ are by hypothesis distinct, our graphs do not contain self-loops.  We will be exclusively concerned with {\bf undirected graphs}, meaning that the edges $(n_1,n_2,\alpha)$ and $(n_2,n_1,\alpha)$ are not distinguished (the edges do not have directional "arrows" associated to them). In Fig.~\ref{fig:graphdefs} we show a pictorial representation of a graph with eight nodes, labelled $n_1$ through $n_8$, and denoted by blue diamonds, black circles, and a red square. This graph also has eight edges. In particular, nodes $n_7$ and $n_8$ are connected by two edges labelled $(n_7,n_8,\alpha)$, with $\alpha\in\{1,\ell_{n_7n_8}=2\}$.

A {\bf partition} of a graph is a subset $P_i\subset N(\mathcal{G})$ of nodes of $\mathcal{G}$ such that for all $p,q\in P_i,\; (p,q,\alpha)\not\in E(\mathcal{G})$. In words, this says that a partition is a subset of nodes such that no two nodes in the subset are connected by edges. Note that in any graph, each node lies in a trivial partition containing itself only (and so the decomposition of a graph into partitions is not unique).

Finally, a {\bf subgraph} $\mathcal{H}\subset\mathcal{G}$\footnote{We will use the subset symbol to denote the graph subgraph relationship. As graphs will always be denoted by calligraphic letters, no confusion should arise.} of a graph $\mathcal{G}$ is a graph such that the set of nodes $N(\mathcal{H})\subset N(\mathcal{G})$, the set of edges $E(\mathcal{H})\subset E(\mathcal{G})$, subject to the restriction that for each edge $(p,q,\alpha)\in E(\mathcal{H})$, we have $p,q\in N(\mathcal{H})$. In words, a subgraph of a graph consists of a subset of nodes, and a subset of edges \emph{connecting those nodes only}. In the graph shown in Fig.~\ref{fig:graphdefs} for example, one subgraph $\mathcal{H}$ has $N(\mathcal{H})=\{n_1,n_2,n_4\}$, $E(\mathcal{H})=\{(n_1,n_4,1),(n_2,n_4,1)\}$. This subgraph {\blue is called $\mathbf{2}${\bf -partite} (or {\bf bipartite}),} since all nodes can be placed into the two partitions $P_1=\{n_1,n_2\}$ and $P_2=\{n_4\}$. {\blue Note also that in general a subgraph containing nodes $n_1$ and $n_2$ need not contain all the edges connecting these two nodes. However, for our purposes, we will only make use of subgraphs $\mathcal{H}\subset\mathcal{G}$ which contain all edges from $E(\mathcal{G})$ connecting nodes in $N(\mathcal{H})$.}  

 We now show that band structures in momentum space map to N-partite graphs, where the $N$ partitions correspond to high-symmetry $\mathbf{k}-vectors$ in the BZ. Groups of bands isolated in energy from all others will map to {\bf connected components} of a graph $\mathcal{G}$ - subgraphs whose points are not connected to other nodes in the graph by any edges. Mathematically,  we define subgraphs $\mathcal{H}\subset\mathcal{G}$ such that for all $h_i\in N(\mathcal{H})$ and all $g_i\in N(\mathcal{G})\setminus N(\mathcal{H})$, $(h_i,g_i,\alpha)\not\in E(\mathcal{G})$. For example, the graph shown in Fig.~\ref{fig:graphdefs} has two connected components $\mathcal{H}_1$ and $\mathcal{H}_2$, where $N(\mathcal{H}_1)=\{n_6,n_7,n_8\}$ and $E(\mathcal{H}_1)=\{(n_6,n_7,1),(n_6,n_8,1),(n_7,n_8,1),(n_7,n_8,2)\}$; $\mathcal{H}_2$ contains all the remaining nodes and edges. We will refer to each connected component of a graph as a \emph{disconnected subgraph}.

\subsection{Mapping to graph theory}
We are now in a position to map the problem of forming global band structures to a graph theory problem. We start with a space group, and a set of little group representations at high-symmetry $\mathbf{k}$-vectors. We will define a \emph{connectivity graph}, such that each valid connectivity graph yields a solution of the compatibility relations. First, we identify all symmetry-distinct $\mathbf{k}$-vector manifolds throughout the BZ. For symmorphic groups this list corresponds with the list of $\mathbf{k}$-vectors found on the Bilbao Crystallographic Server\cite{Bilbao3}. {\blue For non-symmorphic groups, there is the additional subtlety that the little group representation according to which a Bloch function transforms may change upon following an energy band smoothly through a full cycle across the BZ\cite{Herring1942}}. The details of how we identify $\mathbf{k}$-manifolds in the non-symmorphic case can be found in Ref.~\onlinecite{graphdatapaper}.
Each $\mathbf{k}$-manifold will label a partition of our connectivity graph. In each partition $\mathbf{k}_i$ (where the index $i$ runs over different $\mathbf{k}$-manifolds in the BZ), we will place a node labelled by the irreps $\rho^a_{\mathbf{k}_i}$ of the little group $G_{\mathbf{k}_i}$ which appear in the $\kdp$ expansion. We allow edges between nodes in partitions $\mathbf{k}_i$ and $\mathbf{k}_j$ only if either $G_{\mathbf{k}_i}\subset G_{\mathbf{k}_j}$ or $G_{\mathbf{k}_j}\subset G_{\mathbf{k}_i}$; these edges represent the continuation of energy bands from a high-symmetry $\mathbf{k}$-vector to a lower symmetry $\mathbf{k}$-manifold. Finally, we would like to enforce the compatibility relations as a constraint on allowed edges. In particular, given a pair of partitions $\mathbf{k}_i$ and $\mathbf{k_t}$ such that $G_{\mathbf{k}_t}\subset G_{\mathbf{k}_i}$, we look at each node $\rho^a_i$ in the partition $\mathbf{k}_i$. We would like to demand that the restriction,
\begin{equation}
\rho^a_i\downarrow G_{\mathbf{k}_t}\approx\bigoplus_{b=1}^{N}\sigma^b_t
\end{equation}
is respected by the edges connecting $\rho^a_i$ to nodes in the partition $\mathbf{k}_t$.

Formalizing the preceding discussion, we define
\begin{defn}\label{def:graphs}
Given a collection of little group representations, $\mathcal{M}$, (i.e. bands) forming a (physical) band representation for a space group $G$, we construct the {\bf connectivity graph} $C_\mathcal{M}$ as follows: 
{we associate a node, $p^a_{\mathbf{k}_i}\in C_\mathcal{M}$, in the graph to each representation $\rho_{\mathbf{k}_i}^a\in\mathcal{M}$ of the little group $G_{\mathbf{k}_i}$ of {\blue (the closure of)} every high-symmetry manifold (point, line, plane, and volume), $\mathbf{k}_i$.}
If an irrep occurs multiple times in $\mathcal{M}$, there is a separate node for each occurence.

{The degree of each node, $p^a_{\mathbf{k}_i}$, is $P_{\mathbf{k}_i}\cdot \mathrm{dim}(\rho^a_{\mathbf{k}_i})$, where $P_{\mathbf{k}_i}$ is the number of high-symmetry manifolds connected to the point $\mathbf{k}_i$}:  $\mathrm{dim}(\rho^a_{\mathbf{k}_i})$ edges lead to each of these other $\mathbf{k}-manifolds$ in the graph, one for each energy band. 
When the manifold corresponding to $\mathbf{k}_i$ is contained within the manifold corresponding to $\mathbf{k}_j$, as in a high-symmetry point that lies on a high-symmetry line, their little groups satisfy $G_{\mathbf{k}_j} \subset G_{\mathbf{k}_i}$. {For each node $p_{\mathbf{k}_i}^a$, we compute
\begin{equation}
\rho_{\mathbf{k}_i}^a\downarrow G_{\mathbf{k}_j}\approx\bigoplus_{b}\rho_{\mathbf{k}_j}^b.
\end{equation}
We then connect each node $p_{\mathbf{k}_j}^b$ to the node $p_{\mathbf{k}_i}^a$ with $\mathrm{dim}(\rho^b_{\mathbf{k}_j})$ edges.
}
\end{defn} 

We have thus reduced the problem of constructing all globally consistent band structures consistent with space group symmetries to the problem of constructing all valid connectivity graphs. This is an enormous simplification: instead of looking at the whole continuum of $\mathbf{k}$-vectors, we need here only look at the finite set (typically around 20) 
of symmetry distinct $\mathbf{k}$-manifolds. We have {\blue explained how to tabulate} the minimal set of $\mathbf{k}$-manifolds for each of the $230$ space groups in Ref.~\onlinecite{graphdatapaper}. {\blue The data is available through the BANDREP program on the Bilbao Crystallographic Server\cite{progbandrep}.} It is convienient here to make a distinction between \emph{maximal} and non-maximal $\mathbf{k}$-manifolds. We call a $\mathbf{k}$-vector $\mathbf{k}_0$ maximal if the little co-group $\overline{G}_\mathbf{k_0}$ (the point group of the little group $G_\mathbf{k_0}$) of the $\mathbf{k}$-manifold $\{\mathbf{k}_0\}$ containing $\mathbf{k}_0$ is not a proper subgroup of the little co-group of any $\mathbf{k}$-manifold connected to $\{\mathbf{k}_0\}$\cite{graphdatapaper}. While we must consider in our connectivity graphs all partitions corresponding to maximal $\mathbf{k}$-manifolds, it happens that some non-maximal $\mathbf{k}$-manifolds give only redundant constraints on the connectivity graph. For instance, all space groups have a $\mathbf{k}$-manifold labelled $GP$, the general position with lowest possible symmetry. In centrosymmetric crystals with time-reversal symmetry, the little co-group of $GP$ contains the composition of inversion and time-reversal symmetries, while in all other cases it is the trivial group. {\blue In either case, the little group of a generic $\mathbf{k}$-point in $GP$ has} exactly one single-valued, and one double-valued representation. 
Furthermore, in every space group, we have that $G_{GP}\subseteq G_\mathbf{k}$ for \emph{every} $\mathbf{k}$-manifold $\{\mathbf{k}\}$. Thus, every $\mathbf{k}$-manifold is compatible with the general position, and the compatibility relations are entirely trivial. As such, we can consistently remove the manifold $GP$ from our connectivity graphs without loss of generality. 

\begin{figure}[t]
\includegraphics[width=0.2\textwidth]{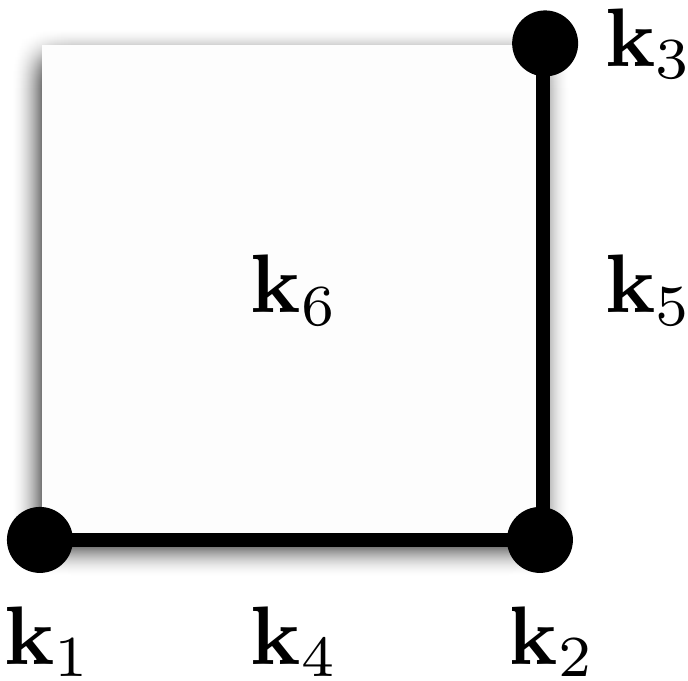}
\caption{Schematic depiction of a triangle-like redundancy. The high symmetry $\mathbf{k}$ points $\mathbf{k}_1$,$\mathbf{k}_2$, and $\mathbf{k}_3$ are connected by the symmetry lines $\mathbf{k}_4$ and $\mathbf{k}_5$. The plane labelled $\mathbf{k}_6$ contains all of these $\mathbf{k}$-manifolds. Because of this, the compatibility relations along any path connecting $\mathbf{k}_1$ and $\mathbf{k}_3$ through $\mathbf{k}_6$ provide no additional symmetry constraints beyond what we get from considering the path $\mathbf{k}_1\rightarrow\mathbf{k}_4\rightarrow\mathbf{k}_2\rightarrow\mathbf{k}_5\rightarrow\mathbf{k}_3$. We can thus safely neglect the manifold $\mathbf{k}_6$ in our construction of compatibility graphs.}\label{fig:triangles}
\end{figure}

A similar redundancy comes from considering closed cycles between compatible $\mathbf{k}$-manifolds, as depicted in Fig.~\ref{fig:triangles}. Suppose that we have six $\mathbf{k}$-manifolds $\mathbf{k}_1,\dots,\mathbf{k}_6$. Suppose also that the little groups of these $\mathbf{k}$ manifolds satisfy the following {\blue group-subgroup} relations:
\begin{align}
G_{\mathbf{k}_6}&\subset G_{\mathbf{k}_i} \forall i \\
G_{\mathbf{k}_5}&\subset G_{\mathbf{k}_3},\;G_{\mathbf{k}_5}\subset G_{\mathbf{k}_2} \\
G_{\mathbf{k}_4}&\subset G_{\mathbf{k}_1},\;G_{\mathbf{k}_4}\subset G_{\mathbf{k}_2}
\end{align} 
Visually in Fig.~\ref{fig:triangles} this has the interpretation that $\mathbf{k}_1,\mathbf{k}_2,$ and $\mathbf{k}_3$ are high-symmetry points. $\mathbf{k}_4$ is the symmetry line connecting $\mathbf{k}_1$ and $\mathbf{k}_2$, and similarly $\mathbf{k}_5$ is the symmetry line connecting $\mathbf{k}_2$ and $\mathbf{k}_3$. Finally, $\mathbf{k}_6$ is the symmetry \emph{plane} containing all of these. In this configuration, it is easy to see that the compatibility relations along $\mathbf{k}_6$ are trivially satisfied if they are satisfied along $\mathbf{k}_4$ and $\mathbf{k}_5$. As such, we can remove the partition corresponding to $\mathbf{k}_6$ from our connectivity graphs. We summarize all such redundancies{\blue , and the algorithms we use to remove them,} for every space group in Ref.~\onlinecite{graphdatapaper}.

With these simplifications in hand, we can now begin to sytematically construct valid connectivity graphs, and hence valid global band structures. The combinatorics involved with directly carrying out this program are still quite daunting, however there are ways to make the task manageable. We present a complete set of algorithms for this task in Ref.~\onlinecite{graphdatapaper}; here we focus on they key conceptual insight -- the reduction to subgraphs.

\subsection{Reduction to subgraphs}

The fundamental building blocks of a connectivity graph are subgraphs consisting of three partitions {\blue corresponding to momenta} $\mathbf{k}_1,\mathbf{k}_2$, and $\mathbf{k}_t$ such that $G_{\mathbf{k}_t}\subset G_{\mathbf{k}_1}$ and $G_{\mathbf{k}_t}\subset G_{\mathbf{k}_2}$. These subgraphs represent, for example,  energy bands at two high-symmetry points $\mathbf{k}_1$ and $\mathbf{k}_2$ connected along a line $\{\mathbf{k}_t\}$. The full connectivity graph is given by a union of permissible subgraphs of this type. As described in Ref.~\onlinecite{graphdatapaper}, by considering every permutation of valid three-partition subgraphs, we can assemble every allowed connectivity graph in this way.

\begin{figure}[t]
\includegraphics[width=0.3\textwidth]{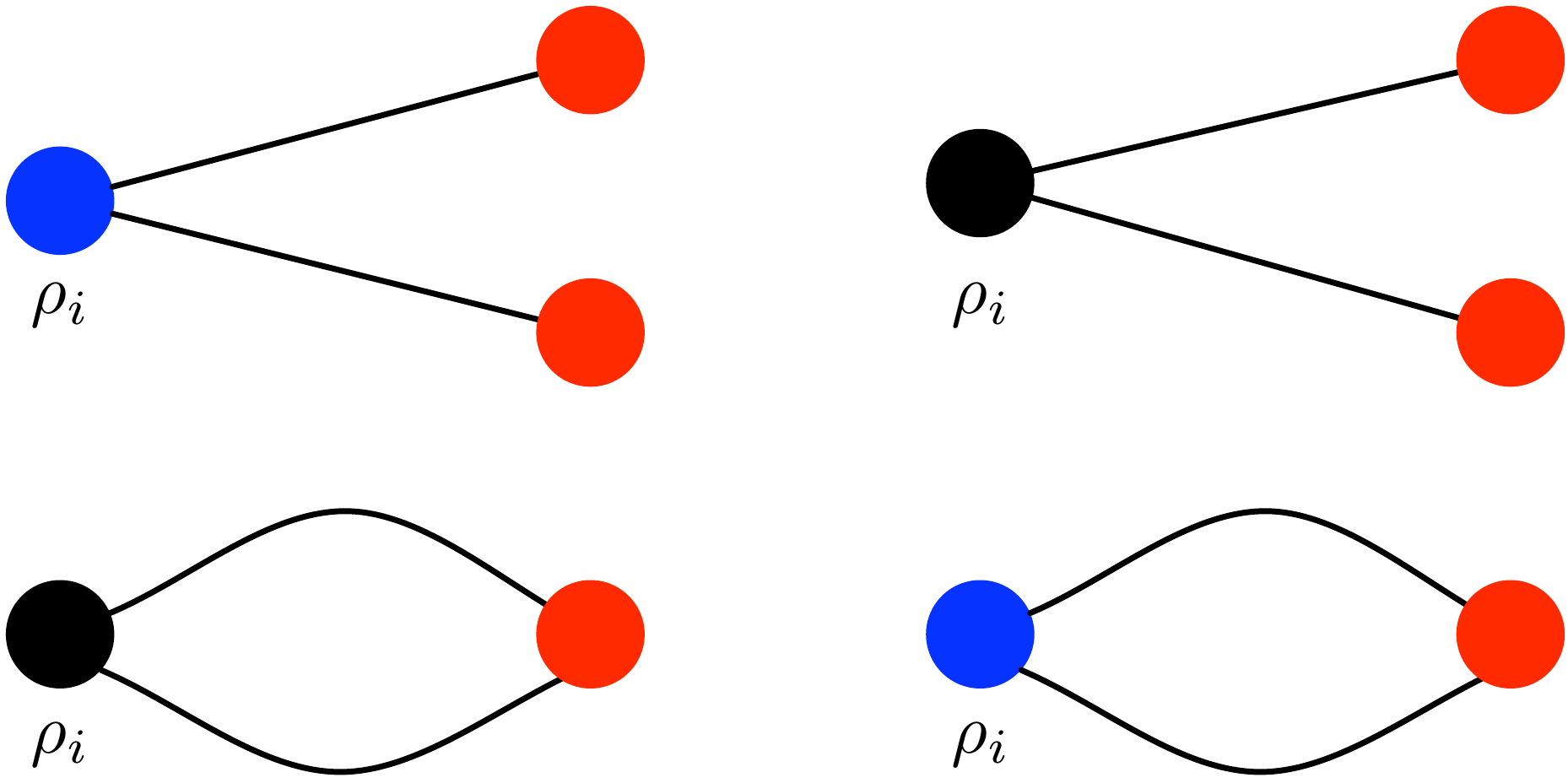}
\caption{The two connectivity subgraphs shown here are trivially isomorphic if the nodes labelled $\rho_i$ correspond to the same little group representations.}\label{fig:permute}
\end{figure}

\begin{figure}[t]
\subfloat[]{
	\includegraphics[height=0.2\textwidth]{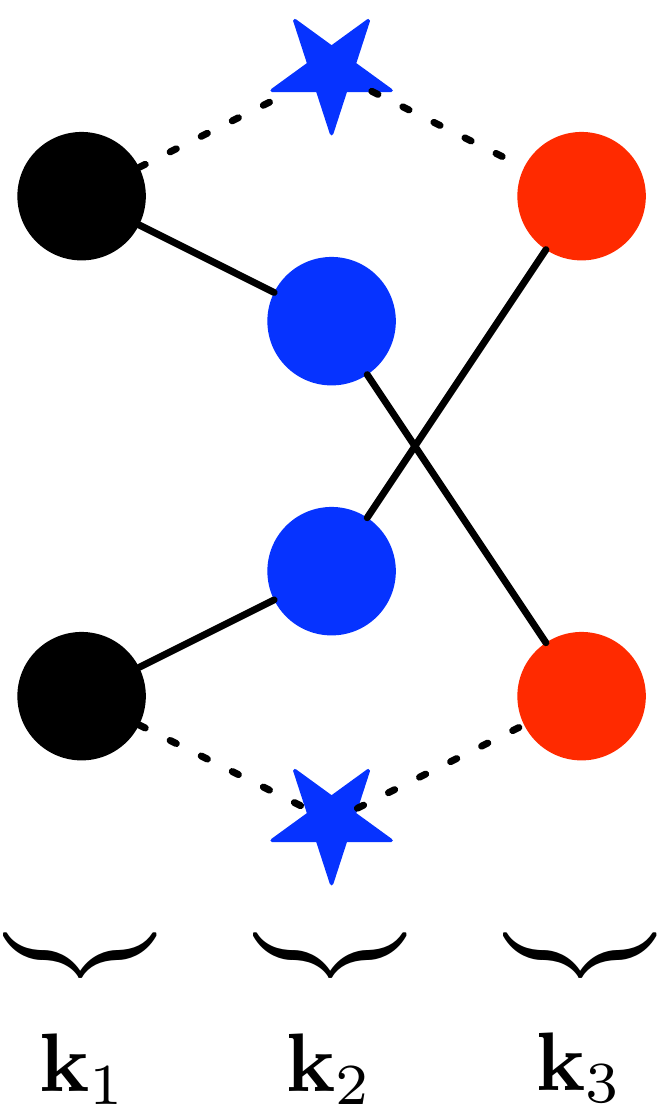}
}\qquad
\subfloat[]{
	\includegraphics[height=0.2\textwidth]{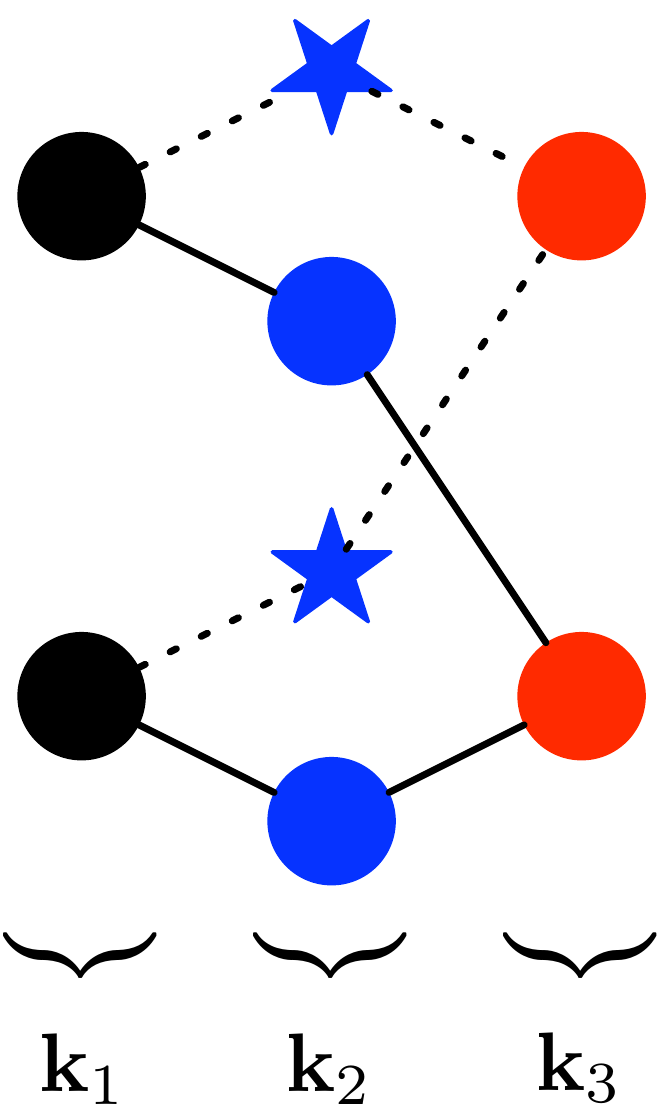}\label{fig:realweyls}
}
\caption{Permissible and non-permissible crossings of edges in connectivity graphs. The blue stars and circles in the $\mathbf{k}_2$ partition correspond to two different irreps of $G_{\mathbf{k}_2}$. (a) shows a non-permissible crossing: when mapped back to a band structure, the energy bands corresponding to the blue circle irreps must cross, and this crossing is not protected. {\blue It will generically either gap or move away from the high-symmetry line.} (b), on the other hand, shows a permissible crossing corresponding to a protected crossing of energy bands carrying different irreps of $G_{\mathbf{k}_2}$.}\label{fig:fakeweyls}
\end{figure}

We must take care, however, that in piecing together the subgraphs we enumerate every connectivity graph only once. As alluded to in Sec.~\ref{sec:globalbs}, when there are multiple nodes $p^{(a)}_{\mathbf{k}_i}$ corresponding to the same representation $\rho_{\mathbf{k}_i}$, then permuting these nodes in a subgraph does not result in a distinct graph, as we illustrate in Fig.~\ref{fig:permute}.
Furthermore, when we account for the fact that the energy bands represented by nodes in our graphs come associated with energy eigenvalues, we arrive at additional constraints from eigenvalue repulsion. Note that the inverse map from a connectivity graph to a band structure requires a certain choice of spatial embedding of the connectivity graph. In particular, all nodes in the partition labelled by $\mathbf{k}_i$ map onto the manifold $\mathbf{k}_i$ in the BZ. When this is done, edges of the connectivity graph may cross, corresponding to crossings of bands in the band structure. Generically, crossings along high-symmetry lines are only protected if the two bands carry different representations of the little group of the line. Accidental crossings of identical representations are not stable to perturbations: they will either gap, {\blue stay gapless over a $1D$ manifold of generic $\mathbf{k}$-vectors (in the case of nodal lines with both inversion and TR symmetry\cite{FangNodalLines})} or in the case of Weyl nodes (which require broken inversion symmetry), they can be pushed away from high-symmetry lines. Because we are interested in classifying generic, stable band structures, we will discount connectivity graphs corresponding to accidental crossings. In Fig.~\ref{fig:fakeweyls} we show examples of a permissible and a non-permissible crossing of bands. In Ref.~\onlinecite{graphdatapaper} we present a set of algorithms that we have developed to systematically rule out unstable crossings. 
\subsection{Example: the cubic crystal revisited}\label{subsec:cubic}

To see some of this machinery at work, we now revisit the $\kdp$ models for the cubic space groups $I432$ ($211$) and $I4_132$ ($214$) which we examined locally in Sec.~\ref{sec:KLexample}. We will now construct the connecivity subgraph for these models along the $\Gamma\leftrightarrow \Lambda\leftrightarrow P$ partitions. 

\begin{figure}[t]
\subfloat[]{
	\includegraphics[width=0.15\textwidth]{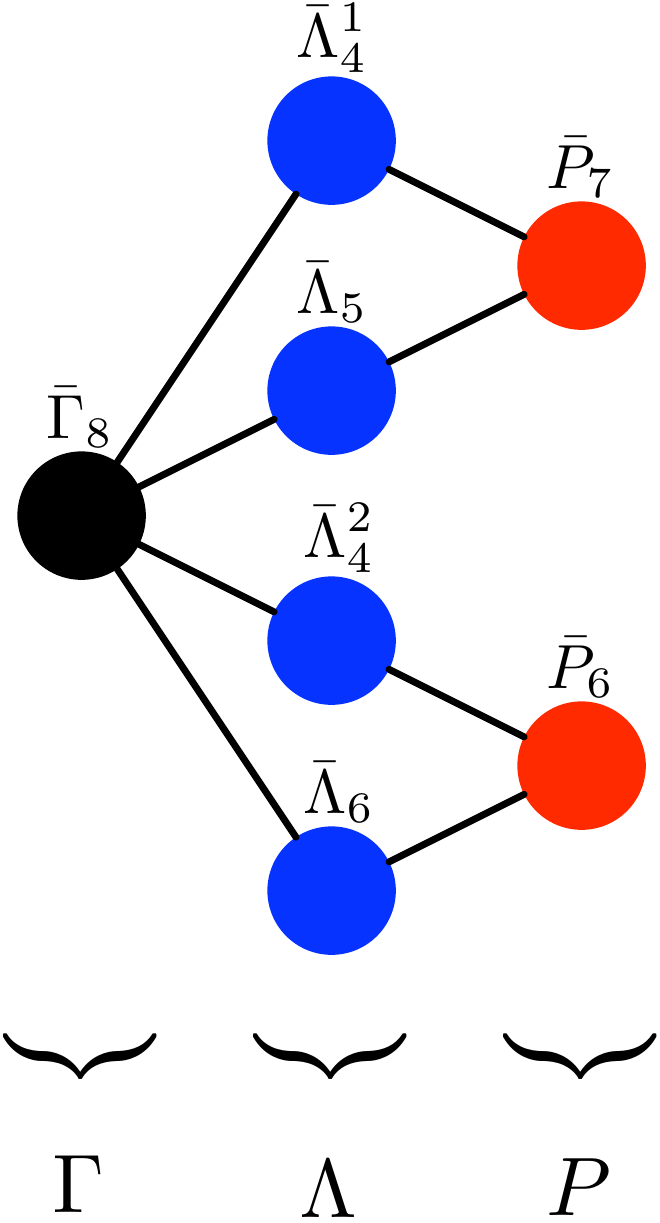}\label{fig:211graph}
}\qquad
\subfloat[]{
	\includegraphics[width=0.15\textwidth]{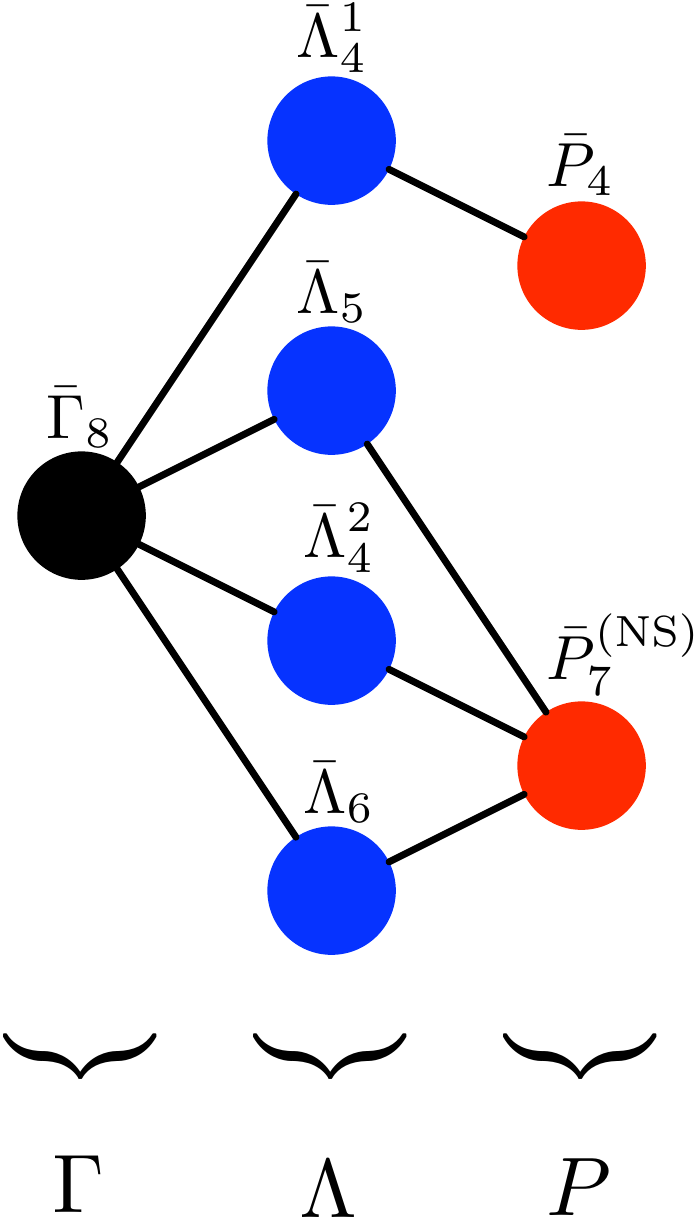}\label{fig:214graph}
}
\caption{Connectivity subgraphs along the $\Gamma\leftrightarrow\Lambda\leftrightarrow P$ line in (a) SG $I432$ ($211$) and (b) $I4_132$ ($214$). In each case there is only one unique subgraph.}
\end{figure}

Let us start with the symmorphic group $I432$ ($211$). In the connectivity subgraph, we have partitions $\Gamma$, $P$, and $\Lambda$. In the $\Gamma$ partition we have a single node corresponding to the four-dimensional $\overline{\Gamma}_8$ little group representation, and in the $P$ partition we have two nodes corresponding to the representations $\overline{P}_6$ and $\overline{P}_7$. We would like to 
now find the compatibility relations for these representations along $\Lambda$. There are three distinct one-dimensional spinful representations distinguished by {\blue their (one-dimensional) matrix value for} $\{C_3|000\}$, and given by
\begin{align}
\overline{\Lambda}_4(\{C_3|000\})&=-1\nonumber\\
\overline{\Lambda}_5(\{C_3|000\})&=e^{-i\pi/3}\nonumber\\
\overline{\Lambda}_6(\{C_3|000\})&=e^{i\pi/3}\label{eq:21xLDreps}
\end{align}
 From the compatibility relations computed in Ref.~\onlinecite{progbandrep}, we have

\begin{align}
\overline{\Gamma}_8\downarrow G_{\Lambda}&\approx\overline{\Lambda}_4\oplus\overline{\Lambda}_4\oplus\overline{\Lambda}_5\oplus\overline{\Lambda}_6\label{eq:211GMLDcomp}
\end{align}
and
\begin{align}
\overline{P}_6\downarrow G_{\Lambda}&\approx\overline{\Lambda}_4\oplus\overline{\Lambda}_6,\nonumber\\
\overline{P}_7\downarrow G_{\Lambda}&\approx\overline{\Lambda}_4\oplus\overline{\Lambda}_5,\label{eq:211PLDcomp}
\end{align}
Thus, in the partition labelled by $\Gamma$, we have a single node $\overline{\Gamma}_8$; in the partition labelled by $P$ we have two nodes $\overline{P}_6$ and $\overline{P}_7$; lastly, in the partition labelled by $\Lambda$ we have four nodes, labelled $\overline{\Lambda}_4^1,\overline{\Lambda}_4^2,\overline{\Lambda}_5,\overline{\Lambda}_6$. Note that the nodes $\overline{\Lambda}_4^1$ and $\overline{\Lambda}_4^2$ correspond to the two copies of the $\overline{\Lambda}_4$ representation appearing in the compatibility relations Eqs.~(\ref{eq:211GMLDcomp}--\ref{eq:211PLDcomp}). We deduce from the compatibility relations that there is a single edge from the node $\overline{\Gamma}_8$ to each of the nodes in the $\Lambda$ partition. Furthermore, there must be an edge connecting the nodes $\overline{\Lambda}_5$ and $\overline{P}_7$, as well as an edge connecting the nodes $\overline{\Lambda}_6$ and $\overline{P}_6$. At first sight, it appears that there are two different ways to connect the $\overline{\Lambda}_4^1$ and $\overline{\Lambda}_4^2$ nodes to the $\overline{P}_6$ and $\overline{P}_7$ nodes: We could either have edges $\{(\overline{\Lambda}_4^1,\overline{P}_6,1),(\overline{\Lambda}_4^2,\overline{P}_7^1,1)\}$, or alternatively $\{(\overline{\Lambda}_4^2,\overline{P}_6,1),(\overline{\Lambda}_4^1,\overline{P}_7^1,1)\}$. However, because the nodes $\overline{\Lambda}_4^i$ correspond to identical irreps of $G_\Lambda$, these two choices of connectivity are isomorphic. We illustrate the full unique connectivity subgraph in Fig.~\ref{fig:211graph}.

Turning next to the nonsymmorphic space group $I4_132$, we find that the construction of the connectivity graph is phenomenologically similar, although the dimensions and labels of the little group irreps are different. Recall from Sec.~\ref{sec:KLexample} that we have a four dimensional representation $\overline{\Gamma}_8$ of $G_\Gamma$, a one-dimensional representation $\overline{P}_4$ of $G_P$, and a three-dimensional representation $\overline{P}^{\mathrm(NS)}_7$ of $G_P$. the little group $G_\Lambda$ of the line $\Lambda$ is the same as in the previous example, and the representations are labelled as in Eqs.~(\ref{eq:21xLDreps}). Consulting the Bilbao Crystallographic Server\cite{progbandrep}, we have the compatibility relations
\begin{align}
\overline{\Gamma}_8\downarrow G_{\Lambda}&\approx\overline{\Lambda}_4\oplus\overline{\Lambda}_4\oplus\overline{\Lambda}_5\oplus\overline{\Lambda}_6 \\
\overline{P}_4\downarrow G_{\Lambda}&\approx\overline{\Lambda}_4\\
\overline{P}^{(\mathrm{NS})}_7\downarrow G_{\Lambda}&\approx\overline{\Lambda}_4\oplus\overline{\Lambda}_5\oplus\overline{\Lambda}_6
\end{align}
We can now construct the connectivity subgraph in analogy with the symmorphic case. We find that the partitions $\Gamma$ and $\Lambda$ are connected identically as in the previous example. Between the $\Lambda$ and $P$ partitions we have edges $\{(\overline{\Lambda}_4^1,\overline{P}_4,1),(\overline{\Lambda}_4^2,\overline{P}_7^{(\mathrm{NS})},1),(\overline{\Lambda}_5,\overline{P}_7^{(\mathrm{NS})},1)
,(\overline{\Lambda}_5,\overline{P}_7^{(\mathrm{NS})},1)\}
$. We draw this graph in Fig.~\ref{fig:214graph}. As in the symmorphic case, there is only one unique subgraph, since the relabelling $\overline{\Lambda}_4^1\leftrightarrow\overline{\Lambda}_4^2$ is unobservable.

\section{Spectral graph theory and connected bands}
One of our main goals in analyzing global band structures is to identify isolated groups of bands which are topologically nontrivial. In our mapping of band structures to graphs given in Sec.~\ref{sec:graphs}, {\blue interconnected groups of bands which can be separated by a gap from any other bands} map to connected components of the connectivity graph. Thus, it will be essential to determine all {\blue connected subgraphs} of a given graph. To this end, we will employ the tools of spectral graph theory. 

\subsection{Spectral graph theory}
To manipulate and analyze graphs, we will require a more compact representation than the pictorial or explicit enumeration of node and edge sets. Let us first order and enumerate the $m$ nodes of $\mathcal{G}$. Using this ordering as a reference basis, we introduce the {\bf adjacency matrix} $A_\mathcal{G}$ of the graph $\mathcal{G}$, an $m\times m$ matrix where the $(ij)$'th entry is the number of edges of the form $(i,j,\alpha)\in E(\mathcal{G})$ connecting the nodes labelled by $i$ and $j$, or zero if no such edges exist. Symbolically, we may write
\begin{equation}\label{eq:adjmatdef}
(A_\mathcal{G})_{ij}=\begin{cases}
\ell_{ij} & \mathrm{if}\;\; (i,j,\ell_{ij})\in E(\mathcal{G}),\;\;\ell_{ij}\geq 1 \\
0 & \mathrm{otherwise}
\end{cases}
\end{equation}
  Because our graphs contain no self-loops, the adjacency matrix is always purely off-diagonal. Furthermore, since our graphs are undirected, the adjacency matrix is symmetric. The sum of the elements in the $i$'th row of the adjacency matrix is known as the {\bf degree} of the node $i$, denoted by
 \begin{equation}\label{eq:degdef}
 d_{i}\equiv \sum_{j=1}^{|N(\mathcal{G})|}(A_\mathcal{G})_{ij},
 \end{equation}
and gives the total number of edges containing the node $n_i$. This allows us to form the {\bf degree matrix}
\begin{equation}\label{eq:degmatdef}
(D_\mathcal{G})_{ij}=d_{i}\delta_{ij}
\end{equation}
For example, the graph $\mathcal{G}$ in Fig.~\ref{fig:graphdefs} has adjacency matrix
\begin{equation}\label{eq:adjmatexample}
A_\mathcal{G}=\;\begin{blockarray}{ccccccccc}
n_1&n_2&n_3&n_8&n_4&n_5&n_6&n_7&\\
\begin{block}{(cccc|ccc|c)c}
0&0&0&0&1&0&0&0& \;n_1\\
0&0&0&0&1&1&0&0& \;n_2\\
0&0&0&0&0&1&0&0& \;n_3\\
0&0&0&0&0&0&1&2& \;n_8\\
\cline{1-8}
1&1&0&0&0&0&0&0& \;n_4\\
0&1&1&0&0&0&0&0& \;n_5\\
0&0&0&1&0&0&0&1& \;n_6\\
\cline{1-8}
0&0&0&2&0&0&1&0& \;n_7\\
\end{block}
\end{blockarray},
\end{equation}
where -- as we will do in all cases where it is not readily apparent -- we have labelled the rows and columns of the matrix by the corresponding node in our chosen ordering. We have ordered the rows and columns of $A_\mathcal{G}$ here to correspond with our $3$-partitioning of the graph $\mathcal{G}$. The horizontal and vertical lines delineate the distinct partitions. This  $3$-partite structure of the graph immediately guarantees that $A_\mathcal{G}$ to be block-off-diagonal, and allows us to build $A_\mathcal{G}$ iteratively by considering the sub-blocks one-by-one. From this adjacency matrix Eq.~(\ref{eq:adjmatexample}) we compute the degree matrix
\begin{equation}\label{eq:degmatexample}
D_\mathcal{G}=\;\begin{blockarray}{ccccccccc}
n_1&n_2&n_3&n_8&n_4&n_5&n_6&n_7&\\
\begin{block}{(cccc|ccc|c)c}
1&0&0&0&0&0&0&0& \;n_1\\
0&2&0&0&0&0&0&0& \;n_2\\
0&0&1&0&0&0&0&0& \;n_3\\
0&0&0&3&0&0&0&0& \;n_8\\
\cline{1-8}
0&0&0&0&2&0&0&0& \;n_4\\
0&0&0&0&0&2&0&0& \;n_5\\
0&0&0&0&0&0&2&0& \;n_6\\
\cline{1-8}
0&0&0&0&0&0&0&3& \;n_7\\
\end{block}
\end{blockarray},
\end{equation}

Next, let us introduce the {\bf Laplacian matrix} for a graph $\mathcal{G}$, defined in terms of the adjacency matrix (\ref{eq:adjmatdef}) and the degree matrix (\ref{eq:degmatdef}) as
\begin{equation}
L_\mathcal{G}\equiv D_\mathcal{G}-A_\mathcal{G}.\label{eq:lapmatdef}
\end{equation}
{\blue As we remind the reader in Appendix~\ref{app:graph}, the zero eigenvectors of the Laplacian $L_\mathcal{G}$ give the connected components of the graph $\mathcal{G}$.} To see this, we note that zero eigenvectors $\vec{f}$ of $L_\mathcal{G}$ satisfy
\begin{align}
(L_\mathcal{G})_{ab}f_b&=0\\ \label{eq:lapprf1}
\sum_{c}(A_\mathcal{G})_{cb}f_a\delta_{ab}-(A_\mathcal{G})_{ab}f_b&=0.
\end{align}
For fixed $a$, we see that this equation is satisfied if $f_b=1$ whenever $(A_{\mathcal{G}})_{ab}\neq 0$. By induction, this allows us to deduce that the vector
\begin{equation}
f_b=\begin{cases}
1, & a\;\mathrm{and}\;b\;\mathrm{lie\;in\;the\;same\;connected\;component} \\
0, &\mathrm{otherwise}
\end{cases}\label{eq:lapprf2}
\end{equation}
is a zero eigenvector of $L_\mathcal{G}$, as supposed. As a corollary, the multiplicity of the zero eigenvalue of $L_\mathcal{G}$ gives the number of connected components of the graph $\mathcal{G}$.

As an example, let us consider the Laplacian matrix for the graph given in Fig.~(\ref{fig:graphdefs}). Subtracting the adjacency matrix Eq.~(\ref{eq:adjmatexample}) from the degree matrix Eq.~(\ref{eq:degmatexample}), we find
\begin{equation}
L_{\mathcal{G}}=\;\begin{blockarray}{ccccccccc}
n_1&n_2&n_3&n_8&n_4&n_5&n_6&n_7&\\
\begin{block}{(cccc|ccc|c)c}
1&0&0&0&-1&0&0&0& \;n_1\\
0&2&0&0&-1&-1&0&0& \;n_2\\
0&0&1&0&0&-1&0&0& \;n_3\\
0&0&0&3&0&0&-1&-2& \;n_8\\
\cline{1-8}
-1&-1&0&0&2&0&0&0& \;n_4\\
0&-1&-1&0&0&2&0&0& \;n_5\\
0&0&0&-1&0&0&2&-1& \;n_6\\
\cline{1-8}
0&0&0&-2&0&0&-1&3& \;n_7\\
\end{block}
\end{blockarray}.
\end{equation}
From Fig.~\ref{fig:graphdefs}, we see that the vectors
\begin{equation}
\mathbf{x}_1^{\mathrm{T}}=\begin{blockarray}{cccccccc}
n_1&n_2&n_3&n_8&n_4&n_5&n_6&n_7\\
\begin{block}{(cccccccc)}
1 & 1 & 1 & 0 & 1 & 1 & 0 & 0\\
\end{block}
\end{blockarray}
\end{equation}
and
\begin{equation}
\mathbf{x}_2^{\mathrm{T}}=\begin{blockarray}{cccccccc}
n_1&n_2&n_3&n_8&n_4&n_5&n_6&n_7\\
\begin{block}{(cccccccc)}
0 & 0 & 0 & 1 & 0 & 0 & 1 & 1\\
\end{block}
\end{blockarray}
\end{equation}
take the value $1$ on exactly one of the connected components of $\mathcal{G}$, and zero on the other connected component. It is straightforward to verify that 
\begin{equation}
L_\mathcal{G}\mathbf{x}_1=L_\mathcal{G}\mathbf{x}_2=0.
\end{equation}
Finally, since the characteristic polynomial $p_{L_\mathcal{G}}(\lambda)$ of the Laplacian matrix is
\begin{equation}
p_{L_\mathcal{G}}(\lambda)=\lambda^2(\lambda-5)(\lambda-3)(\lambda^2-5\lambda+5)(\lambda^2-3\lambda+1),
\end{equation}
we see immediately that there are only two zero-eigenvectors. Thus two vectors $\mathbf{x}_1^T$ and $\mathbf{x}_2^T$ span the entire null space of $L_\mathcal{G}$ in accordance with our claim.

\section{Graphs and band representations}\label{sec:breps}

We can now apply the tools of spectral graph theory to the connectivity graphs of Sec.~\ref{sec:graphs}. This will allow us to identify all of the disconnected components of a conenctivity graph, and hence under the inverse of our graph theory mapping, all of the disconnected groups of bands in a global band structure. As such, we can in principle use these tools to identify all possible insulating band structures allowed in each space group. We are primarily interested, however, in \emph{topological} insulators. To determine whether a given disconnected component of a connectivity graph corresponds to a topological group of bands, we will combine the graph theory mapping presented here with the theory of band representations of Refs.~\onlinecite{NaturePaper,EBRTheoryPaper}. After a brief review of the theory of band representations, we will show how we can enumerate topological band structures by finding the allowed disconnected connectivity graphs corresponding to the so called ``elementary band representations'' (EBRs). We will then see how this works in the example of a two-dimensional inversion symmetric topological insulator, and so demonstrate how our theory contains the standard theory of eigenvalue-based topological invariants\cite{Fu2007,Fang2012,Fu2011}. Finally, we will show how our graph theory method allows us to predict not only topological insulators, but topological semimetals as well.

\subsection{Review of band representations}

Band representations, first introduced by Zak\cite{Zak1982} and applied to spin-orbit coupled and topological materials in a series of papers by the present authors\cite{NaturePaper,EBRTheoryPaper}, relate the real-space localized electronic orbitals in a crystal to the momentum-space band structure. Roughly speaking, orbitals located at a position in the unit cell of a crystal transform under a representation of the symmetry group of the local crystal field (the site-symmetry group) of that position. By applying to this orbital the remaining crystal symmetry operations (including translation), we \emph{induce} (in the sense of representation theory, c.f.~Ref.~\onlinecite{Serre,Fulton2004}) an infinite-dimensional representation of the full space group. The Fourier transform of this representation determines the little group representations at every $\mathbf{k}$-vector of the Brillouin zone in the energy bands coming from these orbitals. For the mathematical details we defer the reader to Refs.~\onlinecite{NaturePaper,EBRTheoryPaper}. 

As constructed, every band structure that can be obtained from localized orbitals that respect all crystal symmetries transforms {\blue according to} some band representation. Furthermore, in analogy with the concept of a finite dimensional irreducible representation, all band representations can be expressed as a sum of ``elementary building bricks.\cite{Zak1980,Zak1981}'' Following Zak we refer to these as \emph{elementary band representations} (EBRs), or, if we also enforce the role of time-reversal symmetry, \emph{physically elementary band representations} (PEBRs). All band structures that can be continuously deformed to an atomic limit without breaking either crystal (or time-reversal) symmetries transform under a sum of (P)EBRs\cite{NaturePaper}. As a corollary, any disconnected group of bands which does not transform under some sum of (P)EBRs must give a topological insulator. In the following, we will show how, by applying our graph theory mapping to the set of bands in a (P)EBR, we are able to enumerate topological band structures.
 
\subsection{Connectivity and topology}

While the theory of band representations tells us which little group representations appear together at all high-symmetry $\mathbf{k}$-points and lines in atomic limit band structures, it does not in itself tell us how these energy bands are allowed to be connected in a real material. Applying the ideas of Ref.~\onlinecite{NaturePaper} discussed above, we see that the connectivity of elementary and physically elementary band representations lies at the heart of the search for topological insulators: if the bands transforming in a (P)EBR are disconnected in the Brillouin zone, then filling only one of the disconnected components will result in a topological insulator (i.e.~a system that cannot be connected to the atomic limit without closing a gap).  While it was originally believed that all elementary band representations (at least without SOC) led to connected bands in momentum space\cite{Michel2001}, we have found this to be an incorrect assumption\cite{NaturePaper,graphdatapaper}. 

Our mapping of global band structures to connectivity graphs is well-suited to tackle the question of connectivity of elementary band representations. Because they arise from atomic-limit systems by construction, the set of little group representations appearing in a (P)EBR \emph{automatically} satisfy all compatibility relations at every $\mathbf{k}$-manifold, and so can be patched together to form at least one consistent, {\blue connected} global band structure; the mapping to connectivity graphs, however, provides an efficient and computationally tractable method to find \emph{all} consistent global band structures for a given (P)EBR, and in particular each disconnected connectivity graph yields a realizable topological insulator (or topological semi-metal, if band crossings are mandated away from high-symmetry points by topological constraints\cite{Hughes2011,Wan11,Yang2014,Kim2015}). By applying the graph theory mapping and spectral graph theory analysis described in Sec.~\ref{sec:graphs}, as well as the practical implementation of these algorithms described in Ref.~\onlinecite{graphdatapaper}, we have enumerated all EBRs and PEBRs that can lead to disconnected bands in the Brillouin zone. There are {\blue 693 such EBRs, and 576 such PEBRs}, and we have tabulated their disconnected connectivity graphs in the BANDREP program on the Bilbao Crystallographic Server.\cite{progbandrep} These represent approximately $10\%$ of the $10403$ total EBRs and PEBRs, leading us to conjecture that at least $10\%$ of all systems should host topologically disconnected bands. {\blue While the Bilbao Crystallographic Server currently gives the a list of nodes in the disconnected components of the connectivity graphs, a graphical depiction will be implemented in the near future\cite{GroupTheoryPaper}.} To see how this works in practice, we will examine below in Sec.~\ref{subsec:checkerboard} two concrete examples of a topological insulator arising from a disconnected connectivity graph, on a checkerboard lattice both with and without inversion symmetry. 	

\subsection{Example: Square Lattice Topological Insulator}\label{subsec:checkerboard}

{\blue Because the simple elementary band representations in the cubic groups we examined in Sec.~\ref{subsec:cubic}, were connected, we shall move on to consider as an example space group $P4mm$ ($99$)}, which has disconnected PEBRs. This is a symmorphic space group with primitive tetragonal Bravais lattice. Taking the lattice vectors $\{\mathbf{e}_1,\mathbf{e}_2,\mathbf{e}_3\}$ to be aligned with the Cartesian directions, we can express the point group $C_{4v}$ as the group generated by
\begin{align}
C_{4z}&:\{\mathbf{e}_1,\mathbf{e}_2,\mathbf{e}_3\}\rightarrow\{-\mathbf{e}_2,\mathbf{e}_1,\mathbf{e}_3\} \\
m_x&:\{\mathbf{e}_1,\mathbf{e}_2,\mathbf{e}_3\}\rightarrow\{-\mathbf{e}_1,\mathbf{e}_2,\mathbf{e}_3\}.
\end{align} 
Since all point group elements act trivially on the $\mathbf{e}_3$ lattice vector, this space group can be viewed as stacks of $2D$ planes, each consisting of a square lattice with wallpaper symmetry group $p4mm$ (wallpaper group number 11, {generated by $C_{4}$ and $m_{x}$}).

\begin{figure}
\includegraphics[width=0.2\textwidth]{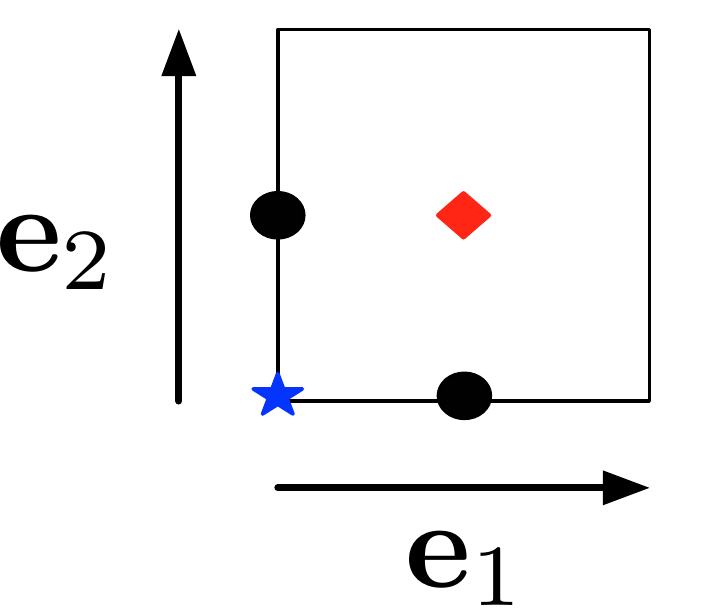}
\caption{Maximal Wyckoff positions for space group $P4mm$ ($99$), {\blue projected along the $\hat{z}=[001]$ direction}. The blue star indicates the $1a$ position at the Bravais lattice sites, the red diamond indicates the $1b$ position at the center of the unit cell, and the black circles denote the $2c$ Wyckoff position at the middle of the {\blue edges}. Because space group $P4mm$ ($99$) has no symmetries constraining $z$, the coordinates of these positions extend in the $z$ direction (perpendicular to the plane of the page)}\label{fig:checker}
\end{figure}

There are three maximal Wyckoff positions {\blue (Wyckoff positions with site-symmetry groups $G_\mathbf{q}$ that are maximal finite subgroups of the space group\cite{Bacry1988,NaturePaper,EBRTheoryPaper})} in this {\blue $3D$} space group, {as indicated in Fig.~\ref{fig:checker}}. First, the $1a$ position corresponds to {vertical lines through} the Bravais lattice sites, and has representative $\{\mathbf{q}^a\}=\{(0,0,z)\}$. Second is the $1b$ position, with representative $\{\mathbf{q}^b\}=\{1/2,1/2,z\}$. This position lies on a vertical line extending upward from the center of the $2D$ square lattice unit cell. Both of these positions have multiplicity $1$, and so have site-symmetry groups isomorphic to the full point group $C_{4v}$. {As the $1a$ position lies on the Bravais lattice sites, its stabilizer $G_{\mathbf{q}^a_1}$ is the full point group, generated by $\{C_{4z}|000\}$ and $\{m_{x}|000\}$. The stabilizer of the $2b$ position, on the other hand, is generated by $\{C_{4z}|100\}$ and $\{m_{x}|100\}$}. In Table~\ref{table:c4v}, we give the character table for $C_{4v}$.
\begin{table}[h]
\begin{tabular}{c|ccccccc}
 Rep & $E$ & $C_{2z}$ & $C_{4z}$ & $\{m, \bar{E}m\}$ & $\{C_{4z}m,\bar{E}C_{4z}m\}$ & $\overline{E}$ & $\overline{E}C_{4z}$\\
 \hline
 $\Gamma_1$ & $1$ & $1$ & $1$ & $1$ & $1$ & $1$ & $1$\\
 $\Gamma_2$ &$1$ &$1$ &$-1$ &$1$ &$-1$&$1$ & $-1$ \\
 $\Gamma_3$ &$1$ &$1$ &$-1$ &$-1$ &$1$&$1$ & $-1$ \\
 $\Gamma_4$ &$1$ &$1$ &$1$ &$-1$ &$-1$ &$1$ & $1$ \\
 $\Gamma_5$ &$2$ &$-2$ &$0$ &$0$ &$0$ &$2$ & $0$ \\
 $\overline{\Gamma}_6$ &$2$ &$0$ &$-\sqrt{2}$ &$0$ &$0$ &$-2$ & $\sqrt{2}$ \\
 $\overline{\Gamma}_7$ &$2$ &$0$ &$\sqrt{2}$ &$0$ &$0$ &$-2$ & $-\sqrt{2}$ \\
 \end{tabular}
 \caption{Character table for the {\blue (double)} point group $C_{4v}$. Note that {\blue the conjucacy class denoted by $\{m,\bar{E}m\}$ contains $m_x$, $m_y$, and their inverses $\bar{E}m_x$ and $\bar{E}m_y$; hence they all have the same characters. Similarly, the conjugacy class $\{C_{4z}m, \bar{E}C_{4z}m\}$ contains $m_{\hat{x}+\hat{y}}$, $m_{\hat{x}-\hat{y}}$, and their inverses.} {Note that spinful orbitals transform in either the $\overline{\Gamma}_6$ or $\overline{\Gamma}_7$ representations depending on their total azimuthal quantum number $m_z$ -- $s$ orbitals (with $m_z=\pm\half$) transform in the $\overline{\Gamma}_7$ representation, as do $P^{\half}$ states, and $P^{\frac{3}{2}}$ states with $m_z=\pm\half$. The remaining $P^{\frac{3}{2}}$ states with $m_z=\pm \frac{3}{2}$ transform in the $\overline{\Gamma}_6$ representation.}\label{table:c4v}}
 \end{table}
Because both the $1a$ and $1b$ Wyckoff positions have multiplicity $1$, all band representations induced from these positions are trivially connected, with connectivity given by the dimension of the site-symmetry representation. To see this, note that since the stabilizer groups of both these positions are isomorphic to the point group, the connectivity graphs for these band representations contain a single node in the partition corresponding to the $\Gamma$ point, and labelled by the site-symmetry irrep. If a connectivity graph has a single node in any partition, then it is impossible to divide it into disconnected subgraphs, {\blue as the node would need to be split in (at least) two}. Thus, all elementary band representations induced from these Wyckoff positions give topologically trivial bands.

\begin{table}[h]
\begin{tabular}{c|ccccc}
 Rep & $E$ & $C_{2z}$ & $\{m, \bar{E}m\}$ & $\{C_{2z}m,\bar{E}C_{2z}m\}$ &$\overline{E}$ \\
 \hline
$\Gamma_1$ & 1 & 1 & 1 & 1 & 1 \\
$\Gamma_2$ & 1 & 1 & -1 & -1 & 1 \\
$\Gamma_3$ & 1 & -1 & -1 & 1 & 1 \\
$\Gamma_4$ & 1 & -1 & 1 & -1 & 1 \\
$\overline{\Gamma}_5$ & 2 & 0 & 0 & 0 & -2
\end{tabular}
\caption{Character table for the {\blue (double)} group $C_{2v}$, for both single- and double-valued representations. The single-valued representations $\Gamma_1$--$\Gamma_4$ are all one dimensional. The unique double-valued representation $\overline{\Gamma}_5$ is the two-dimensional spin-$\half$ representation.  In terms of the Pauli matrices it is given concretely as $\overline{\Gamma}_5(C_{2z})=i\sigma_z,\overline{\Gamma}_5(m)=i\sigma_y$.}
\label{table:c2v}
\end{table}

More interesting for our purposes is the {maximal} $2c$ position, with representatives $\{\mathbf{q}_c^1,\mathbf{q}_c^2\}=\{(0,1/2,z),(1/2,0,z)\}$. The site-symmetry group $G_{\mathbf{q}_1^c}$ of the representative $\mathbf{q}_1^c$ is isomorphic to the group $C_{2v}$, and is generated by $\{C_{2z}|010\}$ and $\{m_x|000\}$. The representations of this group are given in Table~\ref{table:c2v}. Let us focus on the spin-orbit coupled case, where there is {a unique double-valued site-symmetry representation $\overline{\Gamma}_5$, and hence only the band representation $\overline{\Gamma}_5\uparrow G$ can be induced from orbitals at this Wyckoff position.} This representation is generated by a single $s$ orbital with spin up and spin down states at each of the $2c$ sites, forming stacked layers of a checkerboard lattice.  Consulting Refs.~\onlinecite{NaturePaper,progbandrep}, we see that with TR symmetry these orbitals furnish a $4$-band elementary band representation. By computationally constructing the connectivity graphs for this band representation following the algorithms outlined in Sec.~\ref{sec:graphs} and elaborated upon in Ref.~\onlinecite{graphdatapaper}, we find from Ref.~\onlinecite{progbandrep} that this band representation has discconected connectivity graphs {\blue (corresponding to TIs)}, which we now examine further.

We can construct the disconnected realizations of this band representation by examining the compatibility relations for space group $P4mm$ ($99$). The maximal $\mathbf{k}$-vectors in the Brillouin zone relevant to the $2D$ system are $\Gamma=(0,0,0),X=(0,1/2,0),$ and $M=(1/2,1/2,0)$. Note that $C_{4z}$ symmetry relates {\blue $X$} to $ X'=(1/2,0,0)$. {At all these high-symmetry points,} the little groups and their representations are independent of $k_z$. The little co-groups of the points $\Gamma$ and $M$ are the full point group, $C_{4v}$, while the little co-group of {\blue $X$} is isomorphic to the group $C_{2v}$. To examine {$\mathbf{k}$-space} compatibility, we also look at the lines connecting {\blue $\Gamma,\; M,$ and {\blue $X$}}. Particularly relevant are the lines $\Sigma=(k_x,k_x,0)$, $\Delta=(0,k_y,0)$ and $T=(k_x,1/2,0)$, {\blue with $k_x,k_y\in[0,\half]$}. $\Sigma$ connects the points $\Gamma$ and $M$, and has little co-group $C_s$ generated by $m_xC_{4z}$. The line $\Delta$ connects $\Gamma$ and $X$, and has little co-group $C_s$ generated by $m_x$. Finally, the line $T$ connects $X$ and $M$, and has little co-group $C_s$ generated by $m_y$. 
$C_s$ has two double valued representations which we denote by $\rho_\pm$, both one dimensional, and distinguished by whether the mirror is represented by $i$ or $-i$. {\blue Because the group $P4mm$ is symmorphic, the representation $\rho_+$ of $C_s$ uniquely determines the little group representations $\overline{\Sigma}_4$ of $G_\Sigma$ and $\overline{T}_4$ of $G_T$. Similarly, the representation $\rho_-$ uniquely determines the little group representations $\overline{\Sigma}_3$ of $G_\Sigma$ and $\overline{T}_3$ of $G_T$. Physically, this means that the representations of the little groups $G_\Sigma$ and $G_T$ do not change as we move along the lines $\Sigma$ and $T$}.

Performing the induction procedure for the band representation induced from the $2c$ position, and omitting the details (c.f.~Ref.\onlinecite{NaturePaper,GroupTheoryPaper}), {\blue the representations appearing in the Brillouin zone\cite{progbandrep} are presented in Table~\ref{table:99bandrep}}.
\begin{table}[h]
\begin{tabular}{c|c|c|c|c|c}
$\Gamma$ & $\Sigma$ & $M$ & $T$ & {\blue $X$} & $\Delta$ \\
\hline
$\overline{\Gamma}_6\oplus\overline{\Gamma}_7$ & $2\overline{\Sigma}_3\oplus2\overline{\Sigma}_4$ & $\overline{M}_6\oplus\overline{M}_7$ & $2\overline{T}_3\oplus2\overline{T}_4$ & $\overline{Y}_5\oplus\overline{Y}_5$ & $2\overline{\Delta}_3\oplus2\overline{\Delta}_4$ \Tstrut
\end{tabular}
\caption{Little group representations appearing in the double-valued band rep induced from the $2c$ position in SG $P4mm$ ($99$). Note that we use different letters to label the $C_{4v}$ representations at the $\Gamma$ and $M$ points -- regardless of the letter these correspond to the similarly numbered representations in Table~\ref{table:c4v}.}\label{table:99bandrep}
\end{table}

\begin{figure}[t]
\includegraphics[width=0.5\textwidth]{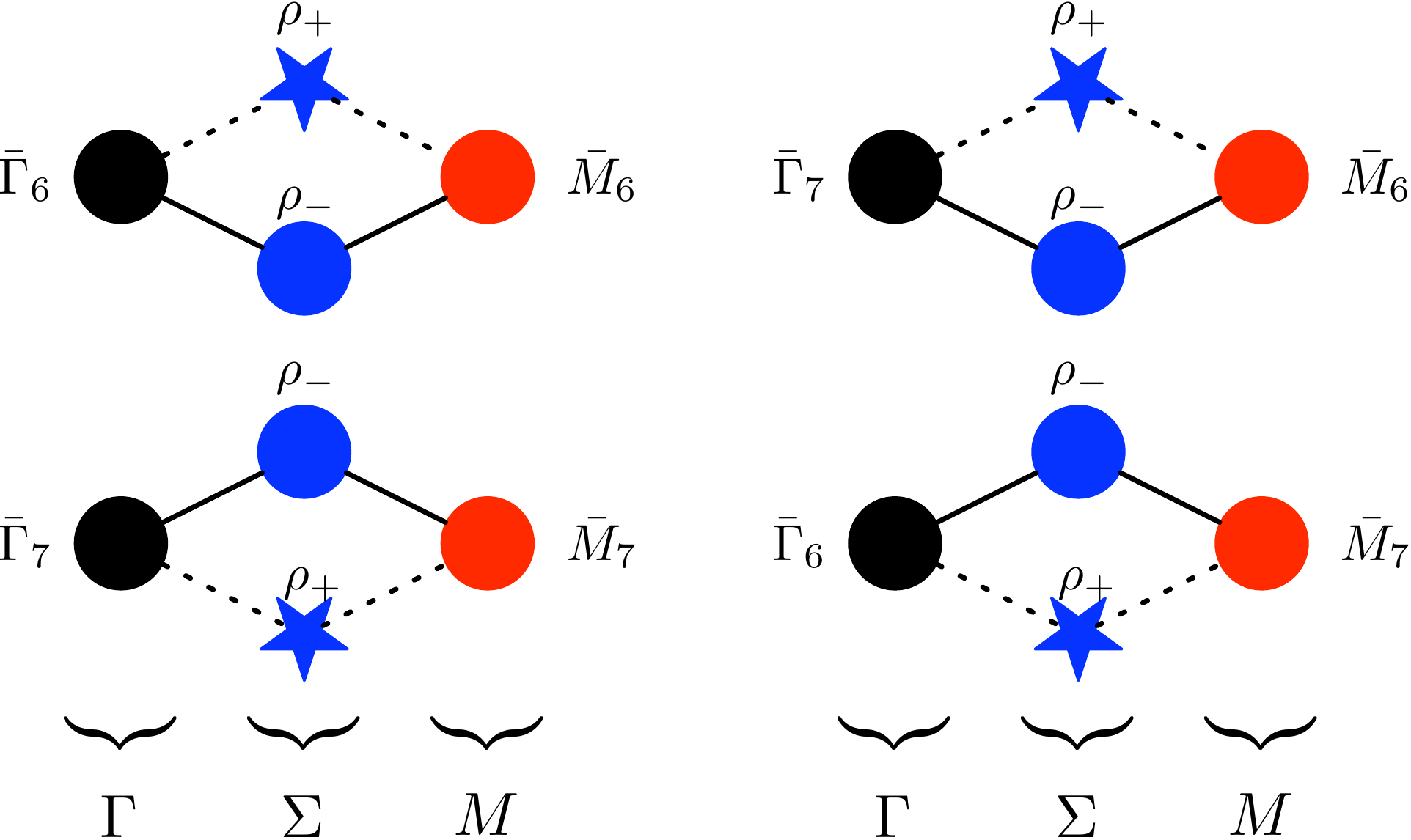}
\caption{Visual depiction of the two different disconnected connectivity graphs for space group $P4mm$ (99), corresponding to the Laplacian matrices in Eqs.~(\ref{eq:lapmatrix1}) and (\ref{eq:lapmatrix2}). In the graph on the left, the $\overline{\Gamma}_6$ little group representation at $\Gamma$ is connected to the $\overline{M}_6$ representation at $M$, while in the graph on the right the $\overline{\Gamma}_7$ little group representation at $\Gamma$ is connected to the $\overline{M}_6$ representation at $M$}\label{fig:sg99graphs}
\end{figure}

Next, we analyze the compatibility between these representations. To do so, note that the little group irreps at $M,\Gamma,$ and {\blue $X$} are all two-dimensional{, as expected from Kramers's theorem}. Furthermore, by consulting Tables~\ref{table:c4v} and \ref{table:c2v}, we see that all mirror elements in these irreps have character 0. This implies that along the high symmetry lines $\Delta,\Sigma,$ and $T$, each two-dimensional little group irrep decomposes into one copy of $\overline{\Sigma}_4$ and one copy of $\overline{\Sigma}_3$, {whose sum of mirror characters is zero}. It is now clear that there are two disconnected compatibility graphs, depending on whether $\overline{\Gamma}_6$ connects to $\overline{M}_6$ or $\overline{M}_7$ along the line $\Sigma$. We show a visual depiction of these two disconnected graphs in Fig.~\ref{fig:sg99graphs}. In Eqs.~(\ref{eq:lapmatrix1}) and (\ref{eq:lapmatrix2}), we show the two distinct Laplacian matrices for the $\Gamma-\Sigma-M$ subgraph of the connectivity graph. Both of these disconnected solutions correspond to topological phases. 
\begin{align}
L_1&=\begin{blockarray}{ccccccccc}
\overline{\Gamma}_6 & \overline{\Gamma}_7 & \overline{M}_6 & \overline{M}_7 & \overline{\Sigma}_4 & \overline{\Sigma}_4 & \overline{\Sigma}_3 &\overline{\Sigma}_3 \\
\begin{block}{(cc|cc|cccc)c}
2&0&0&0&-1&0&-1&0&\;\overline{\Gamma}_6\\
0&2&0&0&0&-1&0&-1&\;\overline{\Gamma}_7\\
\cline{1-8}
0&0&2&0&-1&0&-1&0&\;\overline{M}_6\\
0&0&0&2&0&-1&0&-1&\;\overline{M}_7\\
\cline{1-8}
-1&0&-1&0&2&0&0&0&\;\overline{\Sigma}_4\\
0&-1&0&-1&0&2&0&0&\;\overline{\Sigma}_4\\
-1&0&-1&0&0&0&2&0&\;\overline{\Sigma}_3\\
0&-1&0&-1&0&0&0&2&\;\overline{\Sigma}_3\\
\end{block}
\end{blockarray}\label{eq:lapmatrix1}\\\nonumber\\
L_2&=\begin{blockarray}{ccccccccc}
\overline{\Gamma}_6 & \overline{\Gamma}_7 & \overline{M}_6 & \overline{M}_7 & \overline{\Sigma}_4 & \overline{\Sigma}_4 & \overline{\Sigma}_3 &\overline{\Sigma}_3 \\
\begin{block}{(cc|cc|cccc)c}
2&0&0&0&-1&0&-1&0&\;\overline{\Gamma}_6\\
0&2&0&0&0&-1&0&-1&\;\overline{\Gamma}_7\\
\cline{1-8}
0&0&2&0&0&-1&0&-1&\;\overline{M}_6\\
0&0&0&2&-1&0&-1&0&\;\overline{M}_7\\
\cline{1-8}
-1&0&0&-1&2&0&0&0&\;\overline{\Sigma}_4\\
0&-1&-1&0&0&2&0&0&\;\overline{\Sigma}_4\\
-1&0&0&-1&0&0&2&0&\;\overline{\Sigma}_3\\
0&-1&-1&0&0&0&0&2&\;\overline{\Sigma}_3\\
\end{block}
\end{blockarray}\label{eq:lapmatrix2}
\end{align}
{In order to determine whether this is a strong, weak, or crystalline topological phase, we will} impose inversion symmetry as an additional space group symmetry. This will allow us to use the Fu-Kane inversion eigenvalue formula\cite{Fu2007} to compute topological indices from knowledge of only the valence-band little group representations. Adding inverison takes us from space group $P4mm$ ($99$) to space group $P4/mmm$ ($123$). The addition of spatial inversion forces the $z-$coordinate of our symmetry centers $\{(0,\half,z),(\half,0,z)\}$ to be fixed at either $0$ or $1/2$. We focus on the $z=0$ case for concreteness, since the analysis and results are identical for the $z=1/2$ case. {Note that in SG $P4/mmm$ ($123$), the position $\{(0,\half, 0),(\half,0,0)\}$ is conventially denoted $2e$, and we adopt that notation for the remainder of this section.} In addition to the elements of $C_{2v}$ enumerated above, the site-symmetry group of $\mathbf{q}^f_1=(0,1/2,0)$ now also contains $\{I|010\}$. Thus, the site-symmetry group is isomorphic to $D_{2h}$, whose representations we enumerate below. We note, however, that each representation of $D_{2h}$ is generated by taking a representation of $C_{2v}$ and appending inversion represented as either plus or minus the identity matrix. Since we started with s-orbitals in SG $P4mm$ ($99$), we know that $\{I|010\}$ should be represented by the identity matrix in the relevant representation. Thus, when we add inversion to the disconnected band representation in SG $P4mm$ ($99$), we arrive at the band representation induced from the $\overline{\Gamma_5}$ representation of the site-symmetry group $D_{2h}$ in SG $P4/mmm$ ($123$). Consulting Ref.~\onlinecite{progbandrep}, we see that this band representation is physically elementary, and allows for disconnected connectivity graphs; this means we can add inversion symmetry without changing the band connectivity. To analyze the topological character of {\blue the phases where this band rep is disconnected}, we construct explicitly the $\mathbf{k}$-dependent band representation matrices. Let $s_i$ be a set of Pauli matrices operating in spin space, and let $\sigma_i$ be a set of Pauli matrices operating in sublattice space $\{\mathbf{q}^f_1,\mathbf{q}^f_2\}$. We start by constructing the matrix representative of inversion $\{I|000\}$. First, we note that inversion acts trivially in spin space. Next, using the fact that
\begin{align}
\{I|000\}\mathbf{q}^f_1&=\mathbf{q}^f_1-(0,1,0) \\
\{I|000\}\mathbf{q}^f_2&=\mathbf{q}^f_2-(1,0,0)
\end{align}
{\blue we deduce using the methods of Refs.~\onlinecite{NaturePaper,EBRTheoryPaper} that} we can represent inversion as
\begin{equation}
{\rho^\mathbf{k}(I)=s_0\otimes\left(\begin{array}{cc}
e^{i\mathbf{k}\cdot\mathbf{e}_2} & 0 \\
 0 & e^{i\mathbf{k}\cdot\mathbf{e}_1}
\end{array}\right)}
\end{equation} 
Let us pause to make the following observation: there are 8 inversion-symmetric momenta in the BZ: $\Gamma$, $X$, $X'$, and $M$, along with the points $Z$, $R$, $R'$, and $A$, directly above them in the $z$ direction. Assume now that our band representation ${\overline{\Gamma}_5^{2f}\uparrow G}$ is realized as disconnected, with two valence and two conduction bands. Because of $C_4$ symmetry, the inversion eigenvalues of the valence bands at $X$ and {\blue $X'$} are the same; so are the inversion eigenvalues at $R$ and $R'$. Examining $\rho^\mathbf{k}(I)$, we see that at $\Gamma$ and $Z$ all four inversion eigenvalues are positive, while at $M$ and $A$ all inversion eigenvalues are negative. At $X$ and $R$ two eigenvalues are positive, and two are negative. Using the well-known relationship between inversion eigenvalues and $\mathbb{Z}_2$ topological indices,\cite{Fu2007} we conclude immediately that {\blue all disconnected realizations of this band representation give} a weak topological insulator, with indices $(0;111)$. In Appendix \ref{app:tb} we present an explicit tight-binding model realizing this topological phase. {\blue We emphasize again that the only phases permitted for this band representation are either a connected topological semimetal, or a disconnected topological insulator. This is analogous to the situation present in graphene\cite{Kane04,Soluyanov2011}, and indeed is true of any elementary band representation which has a disconnected connectivity graph.}

\subsection{Filling constraints: insulators and semimetals}

We have in this section been primarily focused with using the method of connectivity graphs to find topological \emph{insulating} band structures. However, the presence of disconnected groups of bands in a band structure is not in itself enough to guarantee that a material is insulating; it must also be true that connected bands are completely filled, i.e.~that (accounting for {\blue possible} charge-transfer from other bands) there is one electron per band per unit cell. When the electron filling is less than the number of bands in an isolated group, a metallic phase naturally results. 

Crucially, we can use our graph theory mapping to assess when a set of $\kdp$ Hamiltonians can be patched together to form a band structure which allows for a topological semimetallic phase. By this we mean a group of bands that, at the appropriate filling, can host a zero-volume point or line-like Fermi surface (possibly in the presence of additional compensated Fermi pockets). While we defer the systematic treatment of semimetallic connectivity to a future work, we can already see some interesting examples in the connectivity graphs we have examined previously. First, let us return to the example of Sec.~\ref{subsec:cubic}, where we examined 
the connectivity of bands originating from $P_{3/2}$ orbitals in the cubic space groups $I432$ ($211$) and $I4_132$ ($214$). Returning to Figs.~\ref{fig:211graph} and \ref{fig:214graph}, we recall that both band structures have only connected connectivity graphs, with a single representation at the $\Gamma$ point, and with two representations at the $P$ point. We see that at half-filling, the band structure corresponding to the connectivity graph Fig.~\ref{fig:211graph} in SG $I432$ can yield a semimetal with a {\blue pointlike Fermi surface at $\Gamma$}; this is in fact a ``Spin-3/2 Weyl'' semimetal first predicted and discussed in Ref.~\onlinecite{Bradlyn2016}. On the contrary, by counting the number of edges in Fig.~\ref{fig:214graph}, we see that a topological semimetal in SG $I4_132$ ($214$) necessarily hosts {\blue Fermi surface features away from the $\Gamma$ point}. In fact, if we consider additionally the partition $N$ corresponding to the $N$ point (with coordinates $\frac{1}{2}\mathbf{g}_3$), we find\cite{progbandrep} that there are two nodes each corresponding to the two-dimensional representation $\overline{N}_5$ of the little group $G_N$. Comparing the connection $\Gamma\leftrightarrow\Sigma\leftrightarrow N$ with the connection $P\leftrightarrow D\leftrightarrow N$ using the BANDREP application, we can decisively rule out the existence of a {\blue semimetal with features only at $\Gamma$ in this connectivity graph: there must be an additional Fermi pocket centered near $P$ or $N$.}

A second example of topological semimetallic behavior is given by connectivity graphs with symmetry-enforced crossings along high-symmetry lines and planes, as in Fig.~\ref{fig:realweyls}. Along mirror planes in mirror symmetric systems, these types of crossings are generically allowed, but not required. However, along glide planes and screw axes in non-symmorphic systems, these crossings are generically required, leading to the ``movable, but not removable'' band crossings first pointed out by Michel and Zak\cite{Michel1999}. At half-filling, band structures in these glide- and screw-symmetric systems will generically yield Weyl and nodal-line semimetals{\blue \cite{Michel1999,Young2012,Steinberg2014}}.
 
\section{Discussion}

We have shown how graph theory can be used to solve the problem of constructing and classifying global band structures consistent with crystal and time-reversal symmetries. While symmetry strongly constrains the allowed connections of energy bands, there are often many consistent global band structures with different numbers of disconnected bands; our graph theory mapping provides the tools necessary to algorithmically enumerate \emph{all} valid band structures. In combination with the theory of elementary band representations of Refs.~\onlinecite{NaturePaper,EBRTheoryPaper}, we demonstrated that the mapping of global band structures to connectivity graphs gives a powerful method for enumerating topological band structures. In the accompanying Ref.~\onlinecite{graphdatapaper}, we show how to practically implement this programme to algorithmically compute all allowed topologically disconnected elementary band representations in all $230$ space groups both with and without time reversal symmetry; the results of these computations are now publicly available on the Bilbao Crystallographic Server\cite{progbandrep}.

While we have discussed the theory of connectivity graphs largely within the context of the electronic structure of time-reversal invariant crystals, the applications of our method are far grander in scope. The methods here can in principle be adapted to the electronic structure of magnetic materials, the dispersion relation of vibrational modes, and spin-wave spectra. Furthermore, we expect connectivity graphs to play a central role in the classification of topological semimetals, and the theory of non-interacting topological phase transitions.

\appendix

\section{Discussion of the Graph Laplacian}\label{app:graph}
In this Appendix, we explore the motivation for the name ``Graph Laplacian.'' Let us consider the graph $\mathcal{G}$ obtained by discretizing $D$-dimensional Euclidean space with a regular hypercubic lattice with lattice constant $a=1$. This is an infinite graph with nodes $N(\mathcal{G})=\{\mathbf{x}^a=(x^a_1,\cdots,x^a_D)\}$ corresponding to points in the lattice, each with degree (coordination number) $2D$. Edges connect nearest-neighbor lattice sites in the usual way. Now, if we have some function $f:\mathbb{R}^D\rightarrow\mathbb{R}$ on space, we can discretize it by restricting to the points $N(\mathcal{G})$. The discretization of the Laplacian $\nabla^2f$ of $f$ at a lattice site $\mathbf{x}_a$ can be written
\begin{equation}
-\nabla^2f(\mathbf{x}_a)\approx -\sum_{\mu=1}^D \left[f(\mathbf{x}_a+\mathbf{e}^\mu)+f(\mathbf{x}_a-\mathbf{e}^\mu)-2f(\mathbf{x}_a)\right],
\end{equation}
where $\mathbf{e}^\mu=(0,\dots,x_\mu=1,\dots,0)$. We note that each term in square brackets is the discretization of the second derivative of $f$ in the $\mu$ direction. However, also note that
\begin{equation}
\sum_{\mu=1}^D2f(\mathbf{x}_a)=2Df(\mathbf{x}_a)=(D_\mathcal{G})_{ab}f(\mathbf{x}_b).
\end{equation} where $(D_\mathcal{G})_{ab} = 2 D \delta_{ab}$
Futhermore, note that the sum
\begin{equation}
\sum_{\mu=1}^D (f(\mathbf{x}_a+\mathbf{e}^\mu)+f(\mathbf{x}_a-\mathbf{e}^\mu))
\end{equation}
is the sum of $f$ evaluated at all nodes adjacent to $\mathbf{x}_a$. Using the fact that the adjacency matrix $A_\mathcal{G}$ as defined in Eq.~(\ref{eq:adjmatdef}) has matrix elements $(A_\mathcal{G})_{ab}$ equal to $1$ for each node $b$ connected to $a$, and zero otherwise, we can rewrite this as
\begin{equation}
\sum_{\mu=1}^D (f(\mathbf{x}_a+\mathbf{e}^\mu)+f(\mathbf{x}_a-\mathbf{e}^\mu))=(A_{\mathcal{G}})_{ab}f(\mathbf{x}_b).
\end{equation}
Putting it all together the discretized Laplacian of f
\begin{equation}
-\nabla^2f(\mathbf{x}_a)\approx (L_\mathcal{G})_{ab}f(\mathbf{x}_b)\label{eq:lapdiscrete}
\end{equation}
coincides with the graph Laplacian. 

Spectral graph theory starts with the study of the eigenvalues of the graph Laplacian. We can motivate the relation between these eigenvalues and the connectivity of the graph using the correspondence Eq.~(\ref{eq:lapdiscrete}). In particular, let us consider the diffusion equation
\begin{equation}
\partial_t f=\nabla^2 f
\end{equation}
on $\mathbb{R}^d$. We know that if $\phi_E(\mathbf{x})$  
is an eigenfunction of $\nabla^2$ with eigenvalue $-E$, then
\begin{equation}
f_E(\mathbf{x},t)=e^{-Et}\phi_E
\end{equation}
solves the diffusion equation. Because the Laplacian is negative definite, all solutions decay to zero at infinite time unless $E=0$. If we consider diffusion on a subspace of our discretized $\mathbb{R}^D$ given by the union of $N$ disconnected sublattices with Neumann boundary conditions, we know that there will be $N$ eigenfunctions of $\nabla^2$ with eigenvalue $E=0$: these are given by the functions which are $1$ on exactly one of the discs, and zero on the others. These give steady states of the diffusion equation, which are constants on each of the $N$ sublattices.

From this fact we may suspect that zero eigenvectors of the graph Laplacian $L_\mathcal{G}$ correspond to connected components of the graph $\mathcal{G}$, and we would be correct; {\blue we proved this in the main text in Eqs.~(\ref{eq:lapprf1}--\ref{eq:lapprf2}).}

\section{Tight-Binding model for the Disconnected EBR in SG $P4/mmm$}\label{app:tb}

\begin{widetext}

\begin{figure}
    \centering
\subfloat[]{
        \includegraphics[width=0.45\textwidth]{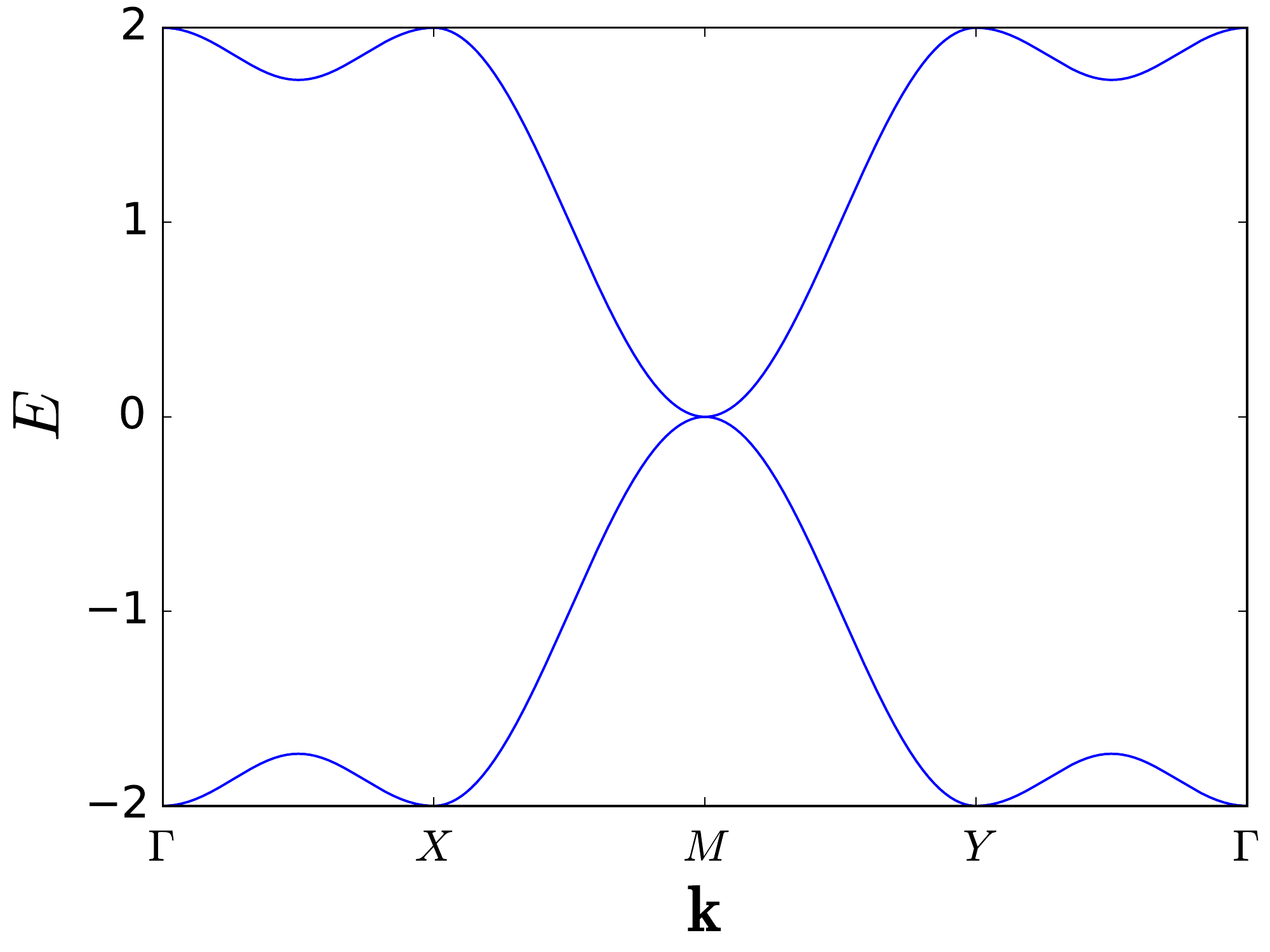}
        \label{fig:123nosoc}
}
\subfloat[]{
    \includegraphics[width=0.45\textwidth]{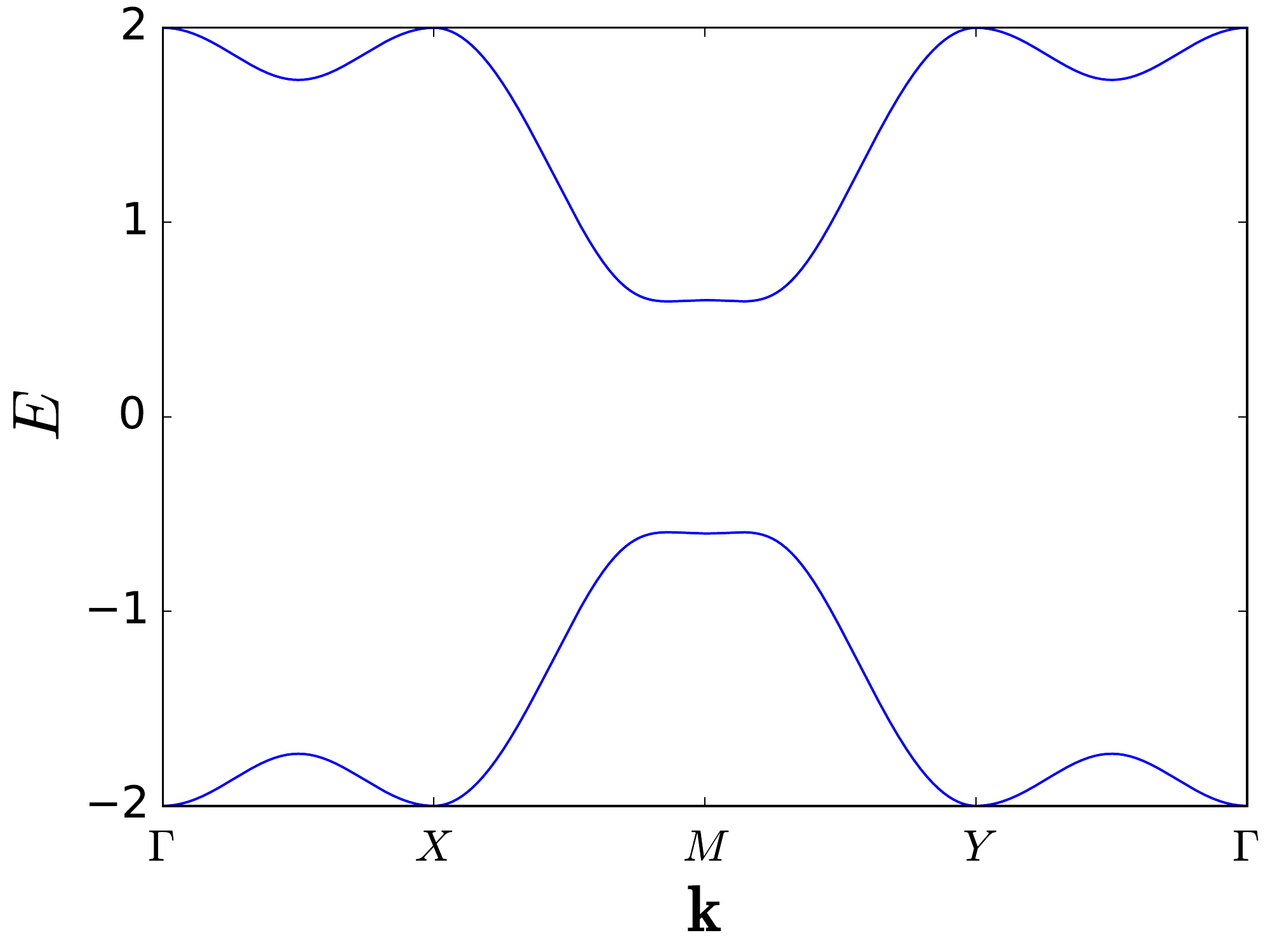}
        \label{fig:123soc}
}
    \caption{Tight-binding band structure for $s$ orbitals at the $2f$ position in SG $P4/mmm$ ($123$) in two dimensions. (a) shows the band structure with no spin-orbit coupling, which yields a semimetal with a gapless point at $M$. (b) shows the spectrum with nonzero SOC. The band structure is fully gapped.}
\end{figure}

\begin{figure}
\centering
\subfloat[]{
        \includegraphics[width=0.5\textwidth]{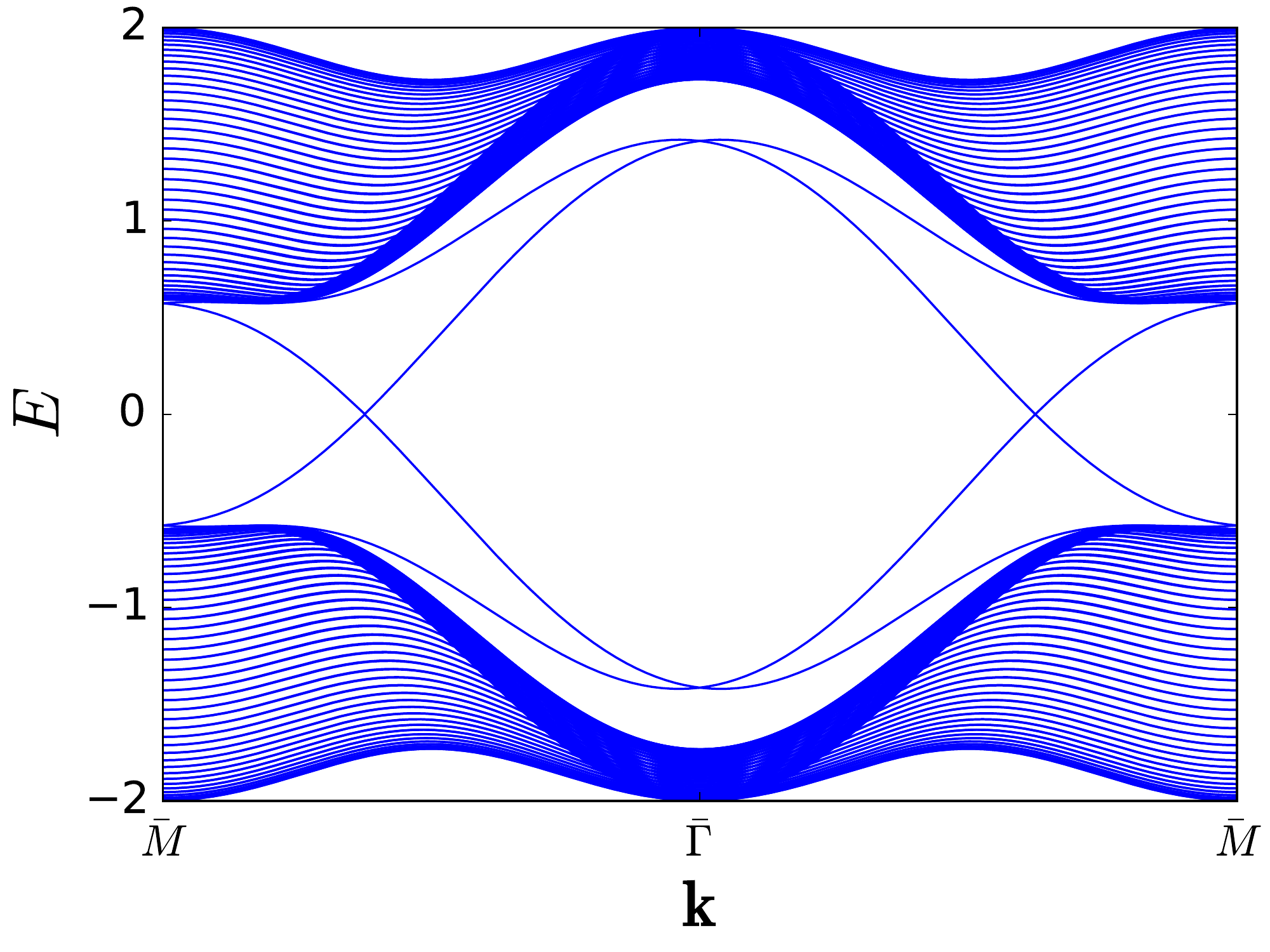}
        \label{fig:123slab}
}
\subfloat[]{
        \includegraphics[width=0.5\textwidth]{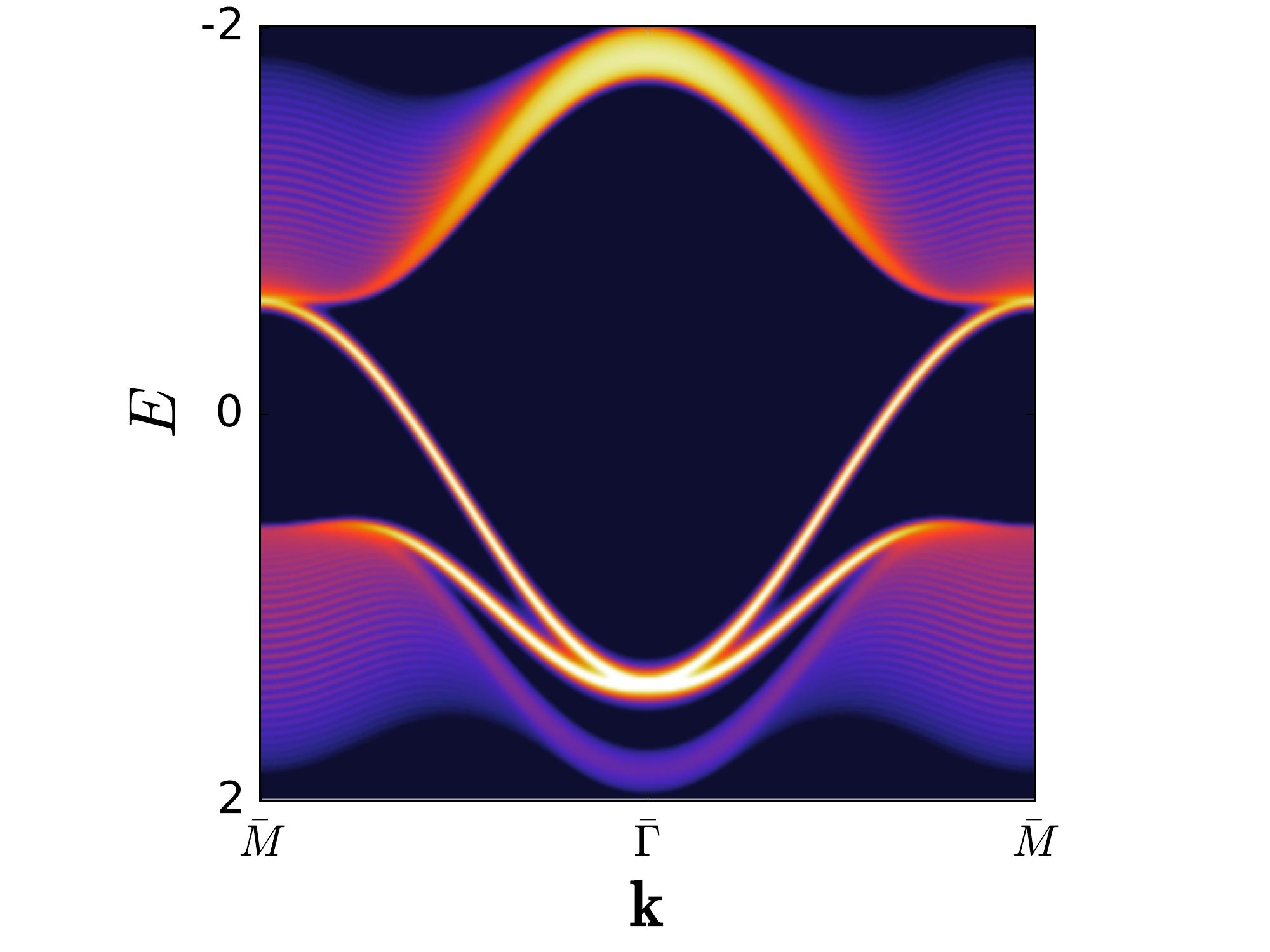}
        \label{fig:123surfspec}
}
     \caption{SG $P4/mmm$ ($123$) tight-binding model on a finite slab {of fifty unit cells}. (a) shows the full slab band structure, with two pairs of counterpropagating edge modes clearly visible in the bulk gap. {Because the total center of charge of orbitals at the $2f$ position are not on Bravais lattice sites, our boundary conditions break inversion symmetry.} (b) shows the surface density of states as a function of energy and surface momentum, showing one pair of counterpropagating edge modes. The other {pair of} edge modes are on the other surface of the slab (not shown).}
     \end{figure}
     
\end{widetext}

It remains for us to show that such a disconnected realization exists. This follows immediately from imposing inversion symmetry on the disconnected band structure we found for SG $P4mm$ ($99$). Inversion forces bands to be doubly degenerate everywhere, {\blue and this leaves our disconnected band structure disconnected}. To see this more clearly, however, we construct an explicit tight-binding model. Noting that in SG $P4/mmm$ ($123$) {along every vertical line in the BZ there is only one allowed  two-dimensional little group representation, we know that the only possible crossings in the $z$ direction can be fourfold crossings of these representations. However, level repulsion ensures that we can eliminate all such crossings. Hence} we can dispense with the $z$-direction in our model. We then construct a $2D$ tight binding model consistent with the space group symmetries -- as with graphene (discussed in Ref.~\onlinecite{Kane04} as well as in the Supplementary Information of Ref.~\onlinecite{NaturePaper}), the effect of inversion in this $2D$ model is to enforce $s_z$ spin conservation. In addition to inversion, we find for the remaining space group generators
\begin{widetext}
\begin{align}
\rho^\mathbf{k}(C_{4z})&=e^{i\mathbf{k}\cdot \mathbf{e}_1/2}e^{-i\pi/4 s_z}\otimes\left(\cos\frac{\mathbf{k}\cdot\mathbf{e}_1}{2}\sigma_x+i\sin\frac{\mathbf{k}\cdot\mathbf{e}_1}{2}\sigma_y\right) \\
\rho^\mathbf{k}(m_x)&=-ie^{i\mathbf{k}\cdot\mathbf{e}_1s_x/2}s_x\otimes\left(\cos\frac{\mathbf{k}\cdot\mathbf{e}_1}{2}\sigma_0-i\sin\frac{\mathbf{k}\cdot\mathbf{e}_1}{2}\sigma_z\right)\\
\rho^\mathbf{k}(T)&=is_y\otimes\sigma_0\mathcal{K},
\end{align}
\end{widetext}
where $\mathcal{K}$ is the complex conjugation operator. We seek a Hamiltonian consistent with these symmetries, that is,
\begin{align}
\rho^{\mathbf{k}}(I)H(\mathbf{k})(\rho^{\mathbf{k}}(I))^{-1}&=H(-\mathbf{k})\\
\rho^{\mathbf{k}}(T)H(\mathbf{k})(\rho^{\mathbf{k}}(T))^{-1}&=H(-\mathbf{k})\\
\rho^{\mathbf{k}}(m_x)H(\mathbf{k})(\rho^{\mathbf{k}}(m_x))^{-1}&=H(m_x\mathbf{k})\\
\rho^{\mathbf{k}}(C_{4z})H(\mathbf{k})(\rho^{\mathbf{k}}(C_{4z}))^{-1}&=H(C_{4z}\mathbf{k})
\end{align}

For the sake of illustration, we will look for the shortest-ranged fully gapped Hamiltonian satisfying these properties. The shortest range Hamiltonian which can suppor a fully gapped insulator reads
\begin{align}
H(\mathbf{k})&=t_1\left[\left(1+\cos\mathbf{k}\cdot\mathbf{e}_1+\cos\mathbf{k}\cdot\mathbf{e}_2+\cos\mathbf{k}\cdot(\mathbf{e}_2-\mathbf{e}_1)\right)\sigma_x\right.\nonumber\\
&\left.+\left(\sin\mathbf{k}\cdot\mathbf{e}_1-\sin\mathbf{k}\cdot\mathbf{e}_2-\sin\mathbf{k}\cdot(\mathbf{e}_2-\mathbf{e}_1)\right)\sigma_y\right]\nonumber \\
&+t_2\left(\cos\mathbf{k}\cdot\mathbf{e}_1-\cos\mathbf{k}\cdot\mathbf{e}_2\right)\sigma_z\nonumber \\
&+\lambda\left[\left(1-\cos\mathbf{k}\cdot\mathbf{e}_1-\cos\mathbf{k}\cdot\mathbf{e}_2+\cos\mathbf{k}\cdot(\mathbf{e}_2-\mathbf{e}_1)\right)s_z\otimes\sigma_y\right.\nonumber\\
&\left.+\left(\sin\mathbf{k}\cdot\mathbf{e}_1-\sin\mathbf{k}\cdot\mathbf{e}_2+\sin\mathbf{k}\cdot(\mathbf{e}_2-\mathbf{e}_1)\right)s_z\otimes\sigma_x\right]
\end{align}
where $t_1$ is a nearest-neighbor inter-sublattice hopping, $t_2$ is a next-nearest-neighbor intra-sublattice hopping, and $\lambda$ is the nearest-neighbor spin-orbit coupling (SOC). We note that a closely related model was recently considered in a different context in Ref.~\onlinecite{Liu2016}. When $\lambda=0$, the system is gapless at the $M$ point, as shown in Figure~\ref{fig:123nosoc}. $s_z$-conserving ``Haldane type'' SOC\cite{haldanemodel,NaturePaper} opens a gap at the $M$ point, giving us a fully disconnected realization of this elementary band representation, as shown in Fig~\ref{fig:123soc}. {Analysis of the inversion eigenvalues reveals that this is a $2D$ strong topological insulator (layers of which form the $3D$ weak topological insulator as above). To see this directly}, we diagonalize our tight-binding model in a slab geometry, periodic in $y$ and with $50$ layers in the $x$-direction. Figure~\ref{fig:123slab} shows the slab band structure as a function of the surface momentum. Two pairs of counterpropagating midgap states are clearly seen, one pair coming from each boundary of the slab. To show that these states are indeed localized to the edge, we compute the surface density of states on the right boundary of the slab, shown in Fig.~\ref{fig:123surfspec}. One pair of counterpropagating edge states is clearly visible. This confirms our assertion that this disconnected elementary band representation realizes a weak $3D$ topological insulator (i.e. a strong $2D$ topological insulator).

\begin{acknowledgements}
{\blue BB would like to thank Ivo Souza, Dustin Ngo, and Ida Momennejad for fruitful discussions. MGV would like to thank Gonzalo Lopez-Garmendia for help with computational work. BB, JC, ZW, and BAB acknowledge the hospitality of the Donostia International Physics Center, where parts of this work were carried out. JC also acknowledges the hospitality of the Kavli Institute for Theoretical Physics, and BAB also acknowledges the hospitality and support of the \'{E}cole Normale Sup\'{e}rieure and Laboratoire de Physique Th\'{e}orique et Hautes Energies. The work of MVG was supported by FIS2016-75862-P and FIS2013-48286-C2-1-P national projects of the Spanish MINECO. The work of LE and MIA was supported by the Government of the Basque Country (project IT779-13)  and the Spanish Ministry of Economy and Competitiveness and FEDER funds (project MAT2015-66441-P). ZW and BAB, as well as part of the development of the initial theory and further ab-initio work, were supported by ARO MURI W911NF-12-1-0461, the Department of Energy de-sc0016239, Simons Investigator Award, the Packard Foundation, and the Schmidt Fund for Innovative Research. The development of the practical part of the theory, tables, some of the code development, and ab-initio work was funded by NSF EAGER Grant No. DMR-1643312, ONR - N00014-14-1-0330, and NSF-MRSEC DMR-1420541. }
\end{acknowledgements}

\bibliography{connectivity}
\end{document}